# Social Norms for Online Communities

Yu Zhang, *Student Member, IEEE,* Jaeok Park, Mihaela van der Schaar, *Fellow, IEEE*
Department of Electrical Engineering, UCLA
yuzhang@ucla.edu, {jaeok, mihaela}@ee.ucla.edu

*Abstract*—*Sustaining cooperation among self-interested agents is critical for the proliferation of emerging online social communities, such as online communities formed through social networking services. Providing incentives for cooperation in social communities is particularly challenging because of their unique features: a large population of anonymous agents interacting infrequently, having asymmetric interests, and dynamically joining and leaving the community; operation errors; and low-cost reputation whitewashing. In this paper, taking these features into consideration, we propose a framework for the design and analysis of a class of incentive schemes based on a social norm, which consists of a reputation scheme and a social strategy. We first define the concept of a sustainable social norm under which every agent has an incentive to follow the social strategy given the reputation scheme. We then formulate the problem of designing an optimal social norm, which selects a social norm that maximizes overall social welfare among all sustainable social norms. Using the proposed framework, we study the structure of optimal social norms and the impacts of punishment lengths and whitewashing on optimal social norms. Our results show that optimal social norms are capable of sustaining cooperation, with the amount of cooperation varying depending on the community characteristics.*

## I. Introduction

Recent developments in technology have expanded the boundaries of communities in which individuals interact with each other. For example, nowadays individuals can obtain valuable information or content from remotely located individuals in a community formed through networking services [1][2][4]. However, a large population and the anonymity of individuals in social communities make it difficult to sustain cooperative behavior among self-interested individuals [3]. For example, it has been reported that so-called free-riding behavior is widely observed in peer-to-peer networks [5][6][7][8]. Hence, incentive schemes are needed to provide individuals with incentives for cooperation.

The literature has proposed various incentive schemes. The popular forms of incentive devices used in many incentive schemes are payment and differential service. Pricing schemes use payments to reward and punish individuals for their behavior. Pricing schemes in principle can lead self-interested individuals to achieve social optimum by internalizing their external effects (see, for example, [9][10]). However, it is often claimed that pricing schemes are impractical because they require an accounting infrastructure [11]. Moreover, the operators of social communities may be reluctant to adopt a pricing scheme when pricing discourages individuals' participation in community activities. Differential service schemes, on the other hand, reward and punish individuals by providing differential services depending on their behavior [12]. Differential services can be provided by community operators or by community members. Community operators can treat individuals differentially (for example, by varying the quality or scope of services) based on the information about the behavior of individuals. Incentive provision by a central entity can offer a robust method to sustain cooperation [13][14]. However, it is impractical in a community with a large population because the burden of a central entity to monitor individuals' behavior and provide differential services for them becomes prohibitively heavy as the number of individuals grows. Alternatively, there are more distributed incentive schemes where community members monitor the behavior of each other and provide differential services based on their observations [15][16][17][18]. Such incentive schemes are based on the principle of reciprocity and can be classified into personal reciprocation (or direct reciprocity) [15][16] and social reciprocation (or indirect reciprocity) [17][18]. In personal reciprocation schemes, individuals can identify each other, and behavior toward an individual is based on their personal experience with the individual. Personal reciprocation is effective in sustaining cooperation in a small community where individuals interact frequently and can identify each other, but it loses its force in a large community where anonymous individuals with asymmetric interests interact infrequently [17]. In social reciprocation schemes, individuals obtain some information about other individuals (for example, rating) and decide their behavior toward an individual based on their information about that individual. Hence, an individual can be rewarded or punished by other individuals in the community who have not had direct interaction with it [17][19]. Since social reciprocation requires neither observable identities nor frequent interactions, it has a potential to form a basis of successful incentive schemes for social communities. As such, this paper is devoted to the study of incentive schemes based on social reciprocation.

Sustaining cooperation using social reciprocation has been investigated in the economics literature using the framework of anonymous random matching games. Social norms have been proposed in [17] and [19] in order to sustain cooperation in a community with a large population of anonymous individuals. In an incentive scheme based on a social norm, each individual is attached a label indicating its reputation, status, etc. which contains information about its past behavior, and individuals with different labels are treated differently by other individuals they interact with. Hence, a social norm can be easily adopted in social communities with an infrastructure that collects, processes, and delivers information about individuals' behavior. However, [17] and [19] have focused on obtaining the Folk Theorem by characterizing the set of equilibrium payoffs that can be achieved by using a strategy based on a social norm when the discount factor is sufficiently close to 1. Our work, on the contrary, addresses the problem of designing a social norm given a discount factor and other parameters arising from practical considerations. Specifically, our work takes into account the following features of social communities:

- *Asymmetry of interests*. As an example, consider a community where individuals with different areas of expertise share knowledge with each other. It will be rarely the case that a pair of individuals has a mutual interest in the expertise of each other. We allow the possibility of asymmetric interests by modeling the interaction between a pair of individuals as a gift-giving game, instead of a prisoner's dilemma game, which assumes mutual interests between a pair of individuals.

- *Report errors*. In an incentive scheme based on a social norm, it is possible that the reputation (or label) of an individual is updated incorrectly because of errors in the report of individuals. Our model incorporates the possibility of report errors, which allows us to analyze its impact on design and performance, whereas most existing works on reputation schemes [15][20][18] adopt an idealized assumption that reputations are always updated correctly.

- *Dynamic change in the population*. The members of a community change over time as individuals gain or lose interest in the services provided by community members. We model this feature by having a constant fraction of individuals leave and join the community in every period. This allows us to study the impact of population turnover on design and performance.

- *Whitewashing reputations.* In an online community, individuals with bad reputations may attempt to whitewash their reputations by joining the community as new members [15]. We consider this possibility and study the design of whitewash-proof social norms and their performance.

The remainder of this paper is organized as follows. In Section II, we describe the repeated matching game and incentive schemes based on a social norm. In Section III, we formulate the problem of designing an optimal social norm. In Section IV, we provide analytical results about optimal social norms. In Section V, we extend our model to address the impacts of variable punishment lengths and whitewashing possibility. We provide simulation results in Section VI, and we conclude the paper in Section VII.

## II. MODEL

### A. Repeated Matching Game

We consider a community where agents can offer a valuable service to other agents. Examples of services are expert knowledge, customer reviews, job information, multimedia files, storage space, and computing power. We consider an infinite-horizon discrete time model with a continuum of agents [15][22]. In a period, each agent generates a service request [27][29], which is sent to another agent that can provide the requested service. We model the request generation and agent selection process by *uniform random matching*, where each agent receives exactly one request in every period, each agent is equally likely to receive the request from a particular agent, and the matching is independent across periods. In a pair of matched agents, the agent that requests a service is called a *client* while the agent that receives a service request is called a *server*. In every period, each agent in the community is involved in two matches, one as a client and the other as a server. Note that the agent with whom an agent interacts as a client can be different from that with which it interacts as a server, reflecting asymmetric interests between a pair of agents.

We model the interaction between a pair of matched agents as a gift-giving game [23]. In a gift-giving game, the server has the binary choice of whether to fulfill or decline the request, while the client has no choice. The server's action determines the payoffs of both agents. If the server fulfils the client's request, the client receives a service benefit of $b > 0$ while the server suffers a service cost of $c > 0$. We assume that $b > c$ so that the service of an agent creates a positive net social benefit. If the server declines the request, both agents receive zero payoffs. The set of actions for the server is denoted by $\mathcal{A} = \{F, D\}$, where $F$ stands for "fulfill" and $D$ for "decline". The payoff matrix of the gift-giving game is presented in Table 1. An agent plays the gift-giving game repeatedly with changing partners until it leaves the community. We assume that at the end of each period a fraction $\alpha \in [0,1]$ of agents in the current population leave and the same amount of new agents join the community. We refer to $\alpha$ as the *turnover rate* [15].

Social welfare in a time period is measured by the average payoff of the agents in that period. Since $b > c$, social welfare is maximized when all the servers choose action $F$ in the gift-giving game they play, which yields payoff $b - c$ to every agent. On the contrary, action $D$ is the dominant strategy for the server in the gift-giving game, which can be considered as the myopic equilibrium of the gift-giving game. When every server chooses its action to maximize its current payoff myopically, an inefficient outcome arises where every agent receives zero payoffs.

TABLE 1. Payoff matrix of a gift-giving game.

|  | Server |  |
|---|---|---|
|  | F | D |
| Client | $b, -c$ | 0, 0 |

*B. Incentive Schemes Based on a Social Norm*

In order to improve the efficiency of the myopic equilibrium, we use incentive schemes based on *social norms*. A social norm is defined as the rules that a group uses to regulate the behavior of members. These rules indicate the established and approved ways of "operating" (e.g. exchanging services) in the group: adherence to these rules is positively rewarded, while failure to follow these rules results in (possibly severe) punishments [26]. This gives social norms a potential to provide incentives for cooperation. We consider a social norm that consists of a *reputation scheme* and a *social strategy*, as in [17] and [19]. A reputation scheme determines the reputations of agents depending on their past actions as a server, while a social strategy prescribes the actions that servers should take depending on the reputations of the matched agents.

Formally, a reputation scheme is represented by three elements $(\Theta, K, \tau)$. $\Theta$ is the set of reputations that an agent can hold, $K \in \Theta$ is the initial reputation attached to newly joining agents, and is the reputation update rule. After a server takes an action, the client sends a report (or feedback) about the action of the server to the third-party device or infrastructure that manages the reputations of agents, but the report is subject to errors with a small probability $\varepsilon$. That is, with probability $\varepsilon$, D is reported when the server takes action F, and vice versa. Assuming a binary set of reports, it is without loss of generality to restrict $\varepsilon$ in $[0, 1/2]$. When $\varepsilon = 1/2$, reports are completely random and do not contain any meaningful information about the actions of servers. We consider a reputation update rule that updates the reputation of a server based only on the reputations of matched agents and the reported action of the server. Then, a reputation update rule can be represented by a mapping $\tau : \Theta \times \Theta \times \mathcal{A} \to \Theta$, where $\tau(\theta, \tilde{\theta}, a_R)$ is the new reputation for a server with current reputation $\theta$ when it is matched with a client with reputation $\tilde{\theta}$ and its action is reported as . A social strategy is represented by a mapping $\sigma : \Theta \times \Theta \to \mathcal{A}$, where $\sigma(\theta, \tilde{\theta})$ is the approved action for a server with reputation $\theta$ that is matched with a client with reputation $\tilde{\theta}$.[1]

To simplify our analysis, we initially impose the following restrictions on reputation schemes we consider.[2]

1) $\Theta$ is a nonempty finite set, i.e., $\Theta = \{0, 1, \ldots, L\}$ for some nonnegative integer $L$.
2) $K = L$.
3) $\tau$ is defined by

---

[1] The strategies in the existing reputation mechanisms [17][19] determine the server's action based solely on the client's reputation, and thus can be considered as a special case of the social strategies proposed in this paper.
[2] We will relax the second and the third restrictions in Section V.

$$\tau\left(\theta,\tilde{\theta},a_R\right) = \begin{cases} \min\{\theta+1, L\} & \text{if } a_R = \sigma\left(\theta,\tilde{\theta}\right), \\ 0 & \text{if } a_R \neq \sigma\left(\theta,\tilde{\theta}\right). \end{cases} \qquad (1)$$

Note that with the above three restrictions a nonnegative integer $L$ completely describes a reputation scheme, and thus a social norm can be represented by a pair $\kappa = (L, \sigma)$. We call the reputation scheme determined by $L$ the *maximum punishment reputation scheme with punishment length* $L$. In the maximum punishment reputation scheme with punishment length $L$, there are $L+1$ reputations, and the initial reputation is specified as $L$. If the reported action of the server is the same as that specified by the social strategy $\sigma$, the server's reputation is increased by 1 while not exceeding $L$. Otherwise, the server's reputation is set as $0$. A schematic representation of a maximum punishment reputation scheme is provided in Fig 1.

Below we summarize the sequence of events in a time period:

1) Agents generate service requests and are matched.
2) Each server observes the reputation of its client and then determines its action.
3) Each client reports the action of its server.
4) The reputations of agents are updated, and each agent observes its new reputation for the next period.
5) A fraction of agents leave the community, and the same amount of new agents join the community.

III. PROBLEM FORMULATION

*A. Stationary Distribution of Reputations*

As time passes, the reputations of agents are updated and agents leave and join the community. Thus, the distribution of reputations evolves over time. Let $\eta^t(\theta)$ be the fraction of $\theta$-agents in the total population at the beginning of an arbitrary period $t$, where a $\theta$-agent means an agent with reputation $\theta$. Suppose that all the agents in the community follow a given social strategy $\sigma$. Then the transition from $\{\eta^t(\theta)\}_{\theta=0}^{L}$ to $\{\eta^{t+1}(\theta)\}_{\theta=0}^{L}$ is determined by the reputation update rule, taking into account the turnover rate $\alpha$ and the error probability $\varepsilon$, as shown in the following expressions:

$$\begin{aligned} \eta^{t+1}(0) &= (1-\alpha)\varepsilon, \\ \eta^{t+1}(\theta) &= (1-\alpha)(1-\varepsilon)\eta^t(\theta-1) \quad \text{for } 1 \leq \theta \leq L-1, \\ \eta^{t+1}(L) &= (1-\alpha)(1-\varepsilon)\{\eta^t(L) + \eta^t(L-1)\} + \alpha. \end{aligned} \qquad (2)$$

Since we are interested in the long-term payoffs of the agents, we study the distribution of reputations in the long run.

*Definition 1 (Stationary distribution)* $\{\eta(\theta)\}$ is a stationary distribution of reputations under the dynamics defined by (2) if it satisfies $\sum_{\theta=0}^{L} \eta(\theta) = 1$ and

$$\eta(0) = (1-\alpha)\varepsilon,$$
$$\eta(\theta) = (1-\alpha)(1-\varepsilon)\eta(\theta-1) \quad \text{for } 1 \leq \theta \leq L-1, \quad (3)$$
$$\eta(L) = (1-\alpha)(1-\varepsilon)\{\eta(L) + \eta(L-1)\} + \alpha.$$

The following lemma shows the existence of and convergence to a unique stationary distribution.

**Lemma 1**. For any $\varepsilon \in [0, 1/2]$ and $\alpha \in [0,1]$, there exists a unique stationary distribution $\{\eta(\theta)\}$ whose expression is given by

$$\eta(\theta) = (1-\alpha)^{\theta+1}(1-\varepsilon)^{\theta}\varepsilon, \text{ for } 0 \leq \theta \leq L-1,$$
$$\eta(L) = \begin{cases} 1 & \text{if } \alpha = \varepsilon = 0, \\ \dfrac{(1-\alpha)^{L+1}(1-\varepsilon)^{L}\varepsilon + \alpha}{1 - (1-\alpha)(1-\varepsilon)} & \text{otherwise.} \end{cases} \quad (4)$$

Moreover, the stationary distribution $\{\eta(\theta)\}$ is reached within $(L+1)$ periods starting from any initial distribution.

*Proof*: Suppose that $\alpha > 0$ or $\varepsilon > 0$. Then (3) has a unique solution

$$\eta(\theta) = (1-\alpha)^{\theta+1}(1-\varepsilon)^{\theta}\varepsilon, \text{ for } 0 \leq \theta \leq L-1,$$
$$\eta(L) = \frac{(1-\alpha)^{L+1}(1-\varepsilon)^{L}\varepsilon + \alpha}{1 - (1-\alpha)(1-\varepsilon)}, \quad (5)$$

which satisfies $\sum_{\theta=0}^{L}\eta(\theta) = 1$. Suppose that $\alpha = 0$ and $\varepsilon = 0$. Then solving (3) together with $\sum_{\theta=0}^{L}\eta(\theta) = 1$ yields a unique solution $\eta(\theta) = 0$ for $0 \leq \theta \leq L-1$ and $\eta(L) = 1$. It is easy to see from the expressions in (2) that $\eta(\theta)$ is reached within $(\theta+1)$ periods, for all $\theta$, starting from any initial distribution. ∎

Since the coefficients in the equations that define a stationary distribution are independent of the social strategy that the agents follow, the stationary distribution is also independent of the social strategy, as can be seen in (4). Thus, we will write the stationary distribution as $\{\eta_L(\theta)\}$ to emphasize its dependence on the reputation scheme, which is represented by $L$.

*B. Sustainable Social Norms*

We now investigate the incentive of agents to follow a prescribed social strategy. For simplicity, we check the incentive of agents at the stationary distribution of reputations, as in [19] and [21]. Since we consider a non-cooperative scenario, we need to check whether an agent can improve its long-term payoff by a unilateral deviation. Note that any unilateral deviation from an individual agent would not affect the evolution of reputations and thus the stationary distribution,[3] because we consider a continuum of agents.

Let $c_\sigma(\theta, \tilde{\theta})$ be the cost suffered by a server with reputation $\theta$ that is matched with a client with reputation $\tilde{\theta}$ and follows a social strategy , i.e., $c_\sigma(\theta,\tilde{\theta}) = c$ if $\sigma(\theta,\tilde{\theta}) = F$ and $c_\sigma(\theta,\tilde{\theta}) = 0$ if $\sigma(\theta,\tilde{\theta}) = D$. Similarly, let $b_\sigma(\theta,\tilde{\theta})$ be the benefit received by a client with reputation $\tilde{\theta}$ that is matched with a server with reputation $\theta$ following a social strategy $\sigma$, i.e., $b_\sigma(\theta,\tilde{\theta}) = b$ if $\sigma(\theta,\tilde{\theta}) = F$ and $b_\sigma(\theta,\tilde{\theta}) = 0$ if $\sigma(\theta,\tilde{\theta}) = D$. Since we consider uniform random matching, the expected period payoff of a $\theta$-agent under social norm $\kappa$ before it is matched is given by

$$v_\kappa(\theta) = \sum_{\tilde{\theta} \in \Theta} \eta_L(\tilde{\theta}) b_\sigma(\tilde{\theta},\theta) - \sum_{\tilde{\theta} \in \Theta} \eta_L(\tilde{\theta}) c_\sigma(\theta,\tilde{\theta}). \tag{6}$$

---
[3] This is true for any deviation by agents of measure zero.

To evaluate the long-term payoff of an agent, we use the discounted sum criterion in which the long-term payoff of an agent is given by the expected value of the sum of discounted period payoffs from the current period. Let $p_\kappa(\theta' \mid \theta)$ be the transition probability that a $\theta$-agent becomes a $\theta'$-agent in the next period under social norm $\kappa$. Since we consider maximum punishment reputation schemes, $p_\kappa(\theta' \mid \theta)$ can be expressed as

$$p_\kappa(\theta' \mid \theta) = \begin{cases} 1 - \varepsilon & \text{if } \theta' = \min\{\theta + 1, L\}, \\ \varepsilon & \text{if } \theta' = 0, \\ 0 & \text{otherwise,} \end{cases} \quad \text{for all } \theta \in \Theta. \tag{7}$$

Then we can compute the long-term payoff of an agent from the current period (before it is matched) by solving the following recursive equations

$$v_\kappa^\infty(\theta) = v_\kappa(\theta) + \delta \sum_{\theta' \in \Theta} p_\kappa(\theta' \mid \theta) v_\kappa^\infty(\theta') \quad \text{for } \theta \in \Theta, \tag{8}$$

where $\delta = \beta(1-\alpha)$ is the weight that an agent puts on its future payoff. Since an agent leaves the community with probability $\alpha$ at the end of the current period, the expected future payoff of a $\theta$-agent is given by $(1-\alpha) \sum_{\theta' \in \Theta} p_\kappa(\theta' \mid \theta) v_\kappa^\infty(\theta')$, assuming that an agent receives zero payoff once it leaves the community. The expected future payoff is multiplied by a common discount factor $\beta \in [0,1)$, which reflects the time preference, or patience, of agents.

Now suppose that an agent deviates and uses a social strategy $\sigma'$ under social norm $\kappa$. Since the deviation of a single agent does not affect the stationary distribution, the expected period payoff of a deviating $\theta$-agent is given by

$$v_{\kappa,\sigma'}(\theta) = \sum_{\tilde{\theta} \in \Theta} \eta_L(\tilde{\theta}) b_\sigma(\tilde{\theta},\theta) + \sum_{\tilde{\theta} \in \Theta} \eta_L(\tilde{\theta}) c_{\sigma'}(\theta,\tilde{\theta}). \tag{9}$$

Let $p_{\kappa,\sigma'}(\theta' \mid \theta,\tilde{\theta})$ be the transition probability that a $\theta$-agent using social strategy $\sigma'$ becomes a $\theta'$-agent in the next period under social norm $\kappa$, when it is matched with a client with reputation $\tilde{\theta}$. For each $\theta$, $\theta' = \min\{\theta+1, L\}$ with probability $(1-\varepsilon)$ and $\theta' = 0$ with probability $\varepsilon$ if $\sigma(\theta,\tilde{\theta}) = \sigma'(\theta,\tilde{\theta})$ while the probabilities are reversed otherwise. Then $p_{\kappa,\sigma'}(\theta'|\theta) = \sum_{\tilde{\theta}\in\Theta} \eta_L(\tilde{\theta}) p_{\kappa,\sigma'}(\theta'|\theta,\tilde{\theta})$ gives the transition probability of a $\theta$-agent before knowing the reputation of its client, and the long-term payoff of a deviating agent from the current period (before it is matched) can be computed by solving

$$v_{\kappa,\sigma'}^{\infty}(\theta) = v_{\kappa,\sigma'}(\theta) + \delta \sum_{\theta'\in\Theta} p_{\kappa,\sigma'}(\theta'|\theta) v_{\kappa,\sigma'}^{\infty}(\theta') \quad \text{for } \theta \in \Theta. \tag{10}$$

In our model, a server decides whether to provide a service or not after it is matched with a client and observes the reputation of the client. Hence, we check the incentive for a server to follow a social strategy at the point when it knows the reputation of the client. Suppose that a server with reputation $\theta$ is matched with a client with reputation $\tilde{\theta}$. When the server follows the social strategy $\sigma$ prescribed by social norm $\kappa$, it receives the long-term payoff $-c_{\sigma}(\theta,\tilde{\theta}) + \delta \sum_{\theta'} p_{\kappa}(\theta'|\theta) v_{\kappa}^{\infty}(\theta')$, excluding the possible benefit as a client. On the contrary, when the server deviates to a social strategy $\sigma'$, it receives the long-term payoff $-c_{\sigma'}(\theta,\tilde{\theta}) + \delta \sum_{\theta'} p_{\kappa,\sigma'}(\theta'|\theta,\tilde{\theta}) v_{\kappa,\sigma'}^{\infty}(\theta')$, again excluding the possible benefit as a client. By comparing these two payoffs, we can check whether a $\theta$-agent has an incentive to deviate to $\sigma'$ when it is matched with a client with reputation $\tilde{\theta}$.

*Definition 2 (Sustainable social norms)* A social norm $\kappa$ is *sustainable* if

$$-c_{\sigma}(\theta,\tilde{\theta}) + \delta \sum_{\theta'} p_{\kappa}(\theta'|\theta) v_{\kappa}^{\infty}(\theta') \geq -c_{\sigma'}(\theta,\tilde{\theta}) + \delta \sum_{\theta'} p_{\kappa,\sigma'}(\theta'|\theta,\tilde{\theta}) v_{\kappa,\sigma'}^{\infty}(\theta') \tag{11}$$

for all $\sigma'$, for all $(\theta,\tilde{\theta})$.

In words, a social norm $\kappa = (L,\sigma)$ is sustainable if no agent can gain from a unilateral deviation regardless of the reputation of the client it is matched with when every other agent follows social strategy $\sigma$ and the reputations are determined by the maximum punishment reputation scheme with punishment length *L*. Thus, under a sustainable social norm, agents follow the prescribed social strategy in their self-interest. Checking whether a social norm is sustainable using the above definition requires computing deviation gains from all possible social strategies, whose computation complexity can be quite high for moderate values of *L*. By employing the criterion of unimprovability in Markov decision theory [28], we establish the one-shot deviation principle for sustainable social norms, which provides simpler conditions. For notation, let  be the cost suffered by a server that takes action $a$, and let $p_{\kappa,a}(\theta'|\theta,\tilde{\theta})$ be the transition probability that a -agent becomes a $\theta'$-agent in the next period under social norm $\kappa$ when it takes action $a$ to a client with reputation $\tilde{\theta}$. The values of $p_{\kappa,a}(\theta'|\theta,\tilde{\theta})$ can be obtained in a similar way to obtain $p_{\kappa,\sigma'}(\theta'|\theta,\tilde{\theta})$, by comparing $a$ with $\sigma(\theta,\tilde{\theta})$.

**Lemma 2 (One-shot Deviation Principle)**. A social norm $\kappa$ is sustainable if and only if

$$c_{\sigma}(\theta,\tilde{\theta}) - c_a \leq \delta \left[ \sum_{\theta'} \{p_{\kappa}(\theta'|\theta) - p_{\kappa,a}(\theta'|\theta,\tilde{\theta})\} v_{\kappa}^{\infty}(\theta') \right] \tag{12}$$

for all $a \in \mathcal{A}$, for all $(\theta,\tilde{\theta})$.

*Proof*: If social norm $\kappa$ is sustainable, then clearly there are no profitable one-shot deviations. We can prove the converse by showing that, if $\kappa$ is not sustainable, there is at least one profitable one-shot deviation. Since $c_\sigma(\theta,\tilde{\theta})$ and $c_a$ are bounded, this is true by the unimprovability property in Markov decision theory [24][25]. ∎

Lemma 2 shows that if an agent cannot gain by unilaterally deviating from $\sigma$ only in the current period and following $\sigma$ afterwards, it cannot gain by switching to any other social strategy $\sigma'$ either, and vice versa. The left-hand side of (12) can be interpreted as the current gain from choosing $a$, while the right-hand side of (12) represents the discounted expected future loss due to the different transition probabilities induced by choosing $a$. Using the one-shot deviation principle, we can derive incentive constraints that characterize sustainable social norms.

First, consider a pair of reputations $(\theta,\tilde{\theta})$ such that $\sigma(\theta,\tilde{\theta}) = F$. If the server with reputation $\theta$ serves the client, it suffers the service cost of $c$ in the current period while its reputation in the next period becomes $\min\{\theta+1, L\}$ with probability $(1-\varepsilon)$ and $0$ with probability $\varepsilon$. Thus, the expected long-term payoff of a $\theta$-agent when it provides a service is given by

$$V_\theta(F;F) = -c + \delta\left[(1-\varepsilon)v_\kappa^\infty\left(\min\{\theta+1,L\}\right) + \varepsilon v_\kappa^\infty(0)\right] \tag{13}$$

On the contrary, if a $\theta$-agent deviates and declines the service request, it avoids the cost of $c$ in the current period while its reputation in the next period becomes $0$ with probability $(1-\varepsilon)$ and $\min\{\theta+1,L\}$ with probability $\varepsilon$. Thus, the expected long-term payoff of a $\theta$-agent when it does not provide a service is given by

$$V_\theta(D;F) = \delta\left[(1-\varepsilon)v_\kappa^\infty(0) + \varepsilon v_\kappa^\infty\left(\min\{\theta+1,L\}\right)\right]. \tag{14}$$

The incentive constraint that a $\theta$-agent does not gain from a one-shot deviation is given by $V_\theta(F;F) \geq V_\theta(D;F)$, which can be expressed as

$$\delta(1-2\varepsilon)\left[v_\kappa^\infty\left(\min\{\theta+1,L\}\right) - v_\kappa^\infty(0)\right] \geq c. \tag{15}$$

Now, consider a pair of reputations $(\theta,\tilde{\theta})$ such that $\sigma(\theta,\tilde{\theta}) = D$. Using a similar argument as above, we can show that the incentive constraint that a $\theta$-agent does not gain from a one-shot deviation can be expressed as

$$\delta(1-2\varepsilon)\left[v_\kappa^\infty\left(\min\{\theta+1,L\}\right) - v_\kappa^\infty(0)\right] \geq -c. \tag{16}$$

Note that (15) implies (16), and thus for $\theta$ such that $\sigma(\theta,\tilde{\theta}) = F$ for some $\tilde{\theta}$, we can check only the first incentive constraint (15). Therefore, a social norm $\kappa$ is sustainable if and only if (15) holds for all $\theta$ such that $\sigma(\theta,\tilde{\theta}) = F$ for some $\tilde{\theta}$ and (16) holds for all $\theta$ such that $\sigma(\theta,\tilde{\theta}) = D$ for all $\tilde{\theta}$. The left-hand side of the incentive constraints (15) and (16) can be interpreted as the loss from punishment that social norm $\kappa$ applies to a $\theta$-agent for not following the social strategy. Therefore, in order to induce a $\theta$-agent to provide a service to some clients, the left-hand side should be at least as large as the service cost $c$, which can be interpreted as the deviation gain. We use $\min_{\theta \in \Theta} \{\delta(1 - 2\varepsilon)[v_\kappa^\infty(\min\{\theta + 1, L\}) - v_\kappa^\infty(0)]\}$ to measure the strength of the *incentive for cooperation* under social norm $\kappa$, where cooperation means providing the requested service in our context.

*C. Social Norm Design Problem*

Since we assume that the community operates at the stationary distribution of reputations, social welfare under social norm $\kappa$ can be computed by

$$U_\kappa = \sum_\theta \eta_L(\theta) v_\kappa(\theta). \tag{17}$$

We assume that the community operator aims to choose a social norm that maximizes social welfare among sustainable social norms. Then the problem of designing a social norm can be formally expressed as

$$\begin{aligned}
\underset{(L,\sigma)}{\text{maximize}} \quad & U_\kappa = \sum_\theta \eta_L(\theta) v_\kappa(\theta) \\
\text{subject to} \quad & \delta(1 - 2\varepsilon)\left[v_\kappa^\infty(\min\{\theta + 1, L\}) - v_\kappa^\infty(0)\right] \geq c, \ \forall \theta \text{ such that } \exists \tilde{\theta} \text{ such that } \sigma(\theta, \tilde{\theta}) = F, \\
& \delta(1 - 2\varepsilon)\left[v_\kappa^\infty(\min\{\theta + 1, L\}) - v_\kappa^\infty(0)\right] \geq -c, \ \forall \theta \text{ such that } \sigma(\theta, \tilde{\theta}) = D \ \forall \tilde{\theta}.
\end{aligned} \tag{18}$$

A social norm that solves the design problem (18) is called an *optimal social norm*.

IV. ANALYSIS OF OPTIMAL SOCIAL NORMS

*A. Optimal Value of the Design Problem*

We first investigate whether there exists a sustainable social norm, i.e., whether the design problem (18) has a feasible solution. Fix the punishment length $L$ and consider a social strategy $\sigma_L^D$ defined by $\sigma_L^D(\theta, \tilde{\theta}) = D$ for all $(\theta, \tilde{\theta})$. Since there is no service provided in the community when all the agents follow $\sigma_L^D$, we have $v_{(L, \sigma_L^D)}^\infty(\theta) = 0$ for all $\theta$. Hence, the relevant incentive constraint (16) is satisfied for all $\theta$, and the social norm $(L, \sigma_L^D)$ is sustainable. This shows that the design problem (18) always has a feasible solution.

Assuming that an optimal social norm exists, let $U^*$ be the optimal value of the design problem (18). In the following proposition, we study the properties of $U^*$.

**Proposition 1.** (i) $0 \leq U^* \leq b - c$.

(ii) $U^* = 0$ if $\dfrac{c}{b} > \dfrac{\beta(1 - \alpha)(1 - 2\varepsilon)}{1 - \beta(1 - \alpha)(2 - 3\varepsilon)}$.

(iii) $U^* \geq [1 - (1 - \alpha)\varepsilon](b - c)$ if $\dfrac{c}{b} \leq \beta(1 - \alpha)(1 - 2\varepsilon)$.

(iv) $U^* < b - c$ if $\varepsilon > 0$.

(v) $U^* = b - c$ if $\varepsilon = 0$ and $\dfrac{c}{b} \leq \beta(1 - \alpha)$.

(vi) $U^* = b - c$ only if $\varepsilon = 0$ and $\dfrac{c}{b} \leq \dfrac{\beta(1 - \alpha)}{1 - \beta(1 - \alpha)}$.

*Proof*: See Appendix A. ∎

We obtain zero social welfare at myopic equilibrium, without using a social norm. Hence, we are interested in whether we can sustain a social norm in which agents cooperate in a positive proportion of matches. In other words, we look for conditions on the parameters $(b, c, \beta, \alpha, \varepsilon)$ that yield $U^* > 0$. From Proposition 1(ii) and (iii), we can regard $c/b \leq [\beta(1-\alpha)(1-2\varepsilon)]/[1-\beta(1-\alpha)(2-3\varepsilon)]$ and $c/b \leq \beta(1-\alpha)(1-2\varepsilon)$ as necessary and sufficient conditions for $U^* > 0$, respectively. Moreover, when there are no report errors (i.e., $\varepsilon = 0$), we can interpret $c/b \leq \beta(1-\alpha)/[1-\beta(1-\alpha)]$ and $c/b \leq \beta(1-\alpha)$ as necessary and sufficient conditions to achieve the maximum social welfare $U^* = b - c$, respectively. As a corollary of Proposition 1, we obtain the following results in the limit.

**Corollary 1.** For any $(b,c)$ such that $b > c$, (i) $U^*$ converges to $b - c$ as $\beta \to 1$, $\alpha \to 0$, and $\varepsilon \to 0$, and (ii) $U^*$ converges to $0$ as $\beta \to 0$, $\alpha \to 1$, or $\varepsilon \to 1/2$. ∎

Corollary 1 shows that we can design a sustainable social norm that achieves near efficiency (i.e., $U^*$ close to $b - c$) when the community conditions are good (i.e., $\beta$ is close to 1, and $\alpha$ and $\varepsilon$ are close to 0). Moreover, it suffices to use only two reputations (i.e., $L = 1$) for the design of such a social norm. On the contrary, no cooperation can be sustained (i.e., $U^* = 0$) when the community conditions are bad (i.e., $\beta$ is close to 0, $\alpha$ is close to 1, or $\varepsilon$ is close to 1/2), as implied by Proposition 1(ii).

*B. Optimal Social Strategies Given a Punishment Length*

In order to obtain analytical results, we consider the design problem (18) with a fixed punishment length $L$, denoted $DP_L$. Note that $DP_L$ has a feasible solution, namely $\sigma_L^D$, for any $L$ and that there are a finite number (total $2^{(L+1)^2}$) of possible social strategies given $L$. Therefore, $DP_L$ has an optimal solution for any $L$. We use $U_L^*$ and $\sigma_L^*$ to denote the optimal value and the optimal social strategy of $DP_L$, respectively. We first show that increasing the punishment length cannot decrease the optimal value.

**Proposition 2.** $U_L^* \geq U_{L'}^*$ for all $L$ and $L'$ such that $L \geq L'$.

*Proof*: See Appendix B. ∎

Proposition 2 shows that $U_L^*$ is non-decreasing in $L$. Since $U_L^* \leq b - c$, we have $U^* = \lim_{L \to \infty} U_L^* = \sup_L U_L^*$. It may be the case that the incentive constraints eventually prevent the optimal value from increasing with $L$ so that the supremum is attained by some finite $L$. If the supremum is not attained, the protocol designer can set an upper bound on $L$ based on practical consideration. Now we analyze the structure of optimal social strategies given a punishment length.

**Proposition 3.** Suppose that $\varepsilon > 0$ and $\alpha < 1$. (i) If $\sigma_L^*(0, \hat{\theta}) = F$ for some $\hat{\theta}$, then $\sigma_L^*(0, \tilde{\theta}) = F$ for all $\tilde{\theta} \geq \min\left\{\ln \frac{c}{b} / \ln \beta, L\right\}$.

(ii) If $\theta \in \{1,...,L-1\}$ satisfies $\theta \geq L - \left(\ln\frac{c}{b} - \ln Y(\alpha,\varepsilon,L)\right)\Big/\ln\beta$, where

$$Y(\alpha,\varepsilon,L) = \frac{(1-\alpha)^{L+1}(1-\varepsilon)^L\varepsilon - (1-\alpha)^{L+2}(1-\varepsilon)^{L+1}\varepsilon}{(1-\alpha)^{L+1}(1-\varepsilon)^L\varepsilon + \alpha} \quad (19)$$

and $\sigma_L^*\left(\theta,\hat{\theta}\right) = F$ for some $\hat{\theta}$, then $\sigma_L^*\left(\theta,L\right) = F$.

(iii) If $\sigma_L^*\left(L,\hat{\theta}\right) = F$ for some $\hat{\theta}$, then $\sigma_L^*\left(L,L\right) = F$.

*Proof*: See Appendix C. ∎

*C. Illustration with L=1 and L=2*

We can represent a social strategy $\sigma_L$ as an $(L+1)\times(L+1)$ matrix whose $(i,j)$-entry is given by $\sigma_L(i-1,j-1)$. Proposition 3 provides some structures of an optimal social strategy $\sigma_L^*$ in the first row and the last column of the matrix representation, but it does not fully characterize the solution of $DP_L$. Here we aim to obtain the solution of $DP_L$ for $L = 1$ and 2 and analyze how it changes with the parameters. We first begin with the case of two reputations, i.e., $L=1$. In this case, if $\sigma_1\left(\theta,\tilde{\theta}\right) = F$ for some $\left(\theta,\tilde{\theta}\right)$, the relevant incentive constraint to sustain $\kappa = (1,\sigma_1)$ is $\delta(1-2\varepsilon)\left[v_\kappa^\infty(1) - v_\kappa^\infty(0)\right] \geq c$. By Proposition 3(i) and (iii), if $\sigma_1^*\left(\theta,\tilde{\theta}\right) = F$ for some $\left(\theta,\tilde{\theta}\right)$, then $\sigma_1^*(0,1) = \sigma_1^*(1,1) = F$, provided that $\varepsilon > 0$ and $\alpha < 1$. Hence, among the total of 16 possible social strategies, only four can be optimal social strategies. These four social strategies are

$$\sigma_1^1 = \begin{bmatrix} D & F \\ F & F \end{bmatrix},\ \sigma_1^2 = \begin{bmatrix} F & F \\ D & F \end{bmatrix},\ \sigma_1^3 = \sigma_1^{D0} = \begin{bmatrix} D & F \\ D & F \end{bmatrix},\ \sigma_1^4 = \sigma_1^D = \begin{bmatrix} D & D \\ D & D \end{bmatrix}. \quad (20)$$

The following proposition specifies the optimal social strategy given the parameters.

**Proposition 4.** Suppose that $0 < (1-\alpha)\varepsilon < 1/2$. Then

$$\sigma_1^* = \begin{cases} \sigma_1^1 & \text{if } 0 < \dfrac{c}{b} \leq \dfrac{\beta(1-\alpha)^2(1-2\varepsilon)\varepsilon}{1+\beta(1-\alpha)^2(1-2\varepsilon)\varepsilon}, \\[6pt] \sigma_1^2 & \text{if } \dfrac{\beta(1-\alpha)^2(1-2\varepsilon)\varepsilon}{1+\beta(1-\alpha)^2(1-2\varepsilon)\varepsilon} < \dfrac{c}{b} \leq \dfrac{\beta(1-\alpha)(1-2\varepsilon)[1-(1-\alpha)\varepsilon]}{1-\beta(1-\alpha)^2(1-2\varepsilon)\varepsilon}, \\[6pt] \sigma_1^3 & \text{if } \dfrac{\beta(1-\alpha)(1-2\varepsilon)[1-(1-\alpha)\varepsilon]}{1-\beta(1-\alpha)^2(1-2\varepsilon)\varepsilon} < \dfrac{c}{b} \leq \beta(1-\alpha)(1-2\varepsilon), \\[6pt] \sigma_1^4 & \text{if } \beta(1-\alpha)(1-2\varepsilon) < \dfrac{c}{b} < 1. \end{cases} \quad (21)$$

*Proof*: Let $\kappa^i = \left(1,\sigma_1^i\right)$, for $i = 1,2,3,4$. We obtain that

$$\begin{aligned} U_{\kappa^1} &= \left(1-\eta_1(0)^2\right)(b-c),\quad U_{\kappa^2} = \left(1-\eta_1(0)\eta_1(1)\right)(b-c), \\ U_{\kappa^3} &= \left(1-\eta_1(0)\right)(b-c),\quad U_{\kappa^4} = 0. \end{aligned} \quad (22)$$

Since $0 < (1-\alpha)\varepsilon < 1/2$, we have $\eta_1(0) < \eta_1(1)$. Thus, we have $U_{\kappa^1} > U_{\kappa^2} > U_{\kappa^3} > U_{\kappa^4}$. Also, we obtain that

$$\begin{aligned}v_{\kappa^1}^\infty(1) - v_{\kappa^1}^\infty(0) = \eta_1(0)(b-c), \quad & v_{\kappa^2}^\infty(1) - v_{\kappa^2}^\infty(0) = b - \eta_1(0)(b-c), \\ v_{\kappa^3}^\infty(1) - v_{\kappa^3}^\infty(0) = b, \quad & v_{\kappa^4}^\infty(1) - v_{\kappa^4}^\infty(0) = 0.\end{aligned} \quad (23)$$

Thus, we have $v_{\kappa^3}^\infty(1) - v_{\kappa^3}^\infty(0) > v_{\kappa^2}^\infty(1) - v_{\kappa^2}^\infty(0) > v_{\kappa^1}^\infty(1) - v_{\kappa^1}^\infty(0) > v_{\kappa^4}^\infty(1) - v_{\kappa^4}^\infty(0)$. By choosing the social strategy that yields the highest social welfare among feasible ones, we obtain the result. ∎

Proposition 4 shows that the optimal social strategy is determined by the ratio of the service cost and benefit, $c/b$. When $c/b$ is sufficiently small, the social strategy $\sigma_1^1$ can be sustained, yielding the highest social welfare among the four candidate social strategies. As $c/b$ increases, the optimal social strategy changes from $\sigma_1^1$ to $\sigma_1^2$ to $\sigma_1^3$ and eventually to $\sigma_1^4$. Fig 2 shows the optimal social strategies with $L=1$ as $c$ varies. The parameters we use to obtain the results in the figures of this paper are set as follows unless otherwise stated: $\beta = 0.8$, $\alpha = 0.1$, $\varepsilon = 0.2$, and $b = 10$. Fig 2(a) plots the incentive for cooperation of the four social strategies. We can find the region of $c$ in which each strategy is sustained by comparing the incentive for cooperation with the service cost $c$ for $\sigma_1^1$, $\sigma_1^2$, and $\sigma_1^3$, and with $-c$ for $\sigma_1^4$. The solid portion of the lines indicates that the strategy is sustained while the dashed portion indicates that the strategy is not sustained. Fig 2(b) plots the social welfare of the four candidate strategies, with solid and dashed portions having the same meanings. The triangle-marked line represents the optimal value, which takes the maximum of the social welfare of all sustained strategies.

Next, we analyze the case of three reputations, i.e. $L=2$. In order to provide a partial characterization of the optimal social strategy $\sigma_2^*$, we introduce the following notation. Let $\sigma_2^\#$ be the social strategy with $L=2$ that maximizes $\min\{v_\kappa^\infty(1) - v_\kappa^\infty(0), v_\kappa^\infty(2) - v_\kappa^\infty(0)\}$ among all the social strategies with $L=2$. Let $\Gamma_2^+$ be the set of all social strategies that satisfy the incentive constraints of $DP_2$ for some parameters $(b, c, \beta, \alpha, \varepsilon)$ satisfying $\varepsilon > 0$, $\alpha < 1$, and (24) below with $\gamma = \delta(1-\varepsilon)$ as defined in Appendix A. Let $\sigma_2^+$ be the social strategy that maximizes social welfare $U_\kappa$ among all the social strategies in $\Gamma_2^+$. Lastly, define a social strategy $\sigma_L^B$ by $\sigma_L^B(L-1, 0) = D$ and $\sigma_L^B(\theta, \tilde{\theta}) = F$ for all $(\theta, \tilde{\theta}) \neq (L-1, 0)$.

**Proposition 5.** Suppose that $\varepsilon > 0$, $\alpha < 1$, and

$$\frac{c}{b} < \frac{\eta_2(2)}{\eta_2(1)} \frac{1-\gamma}{\gamma} < \frac{b}{c}. \quad (24)$$

(i) $\sigma_2^\# = \sigma_2^{D0}$. (ii) If $\eta_2(0) < \eta_2(2)$, then $\sigma_2^+ = \sigma_2^B$.

*Proof*: (i) Let $\kappa = (2, \sigma_2^{D0})$. Then $v_\kappa^\infty(1) - v_\kappa^\infty(0) = v_\kappa^\infty(2) - v_\kappa^\infty(0) = b$. We can show that, under the given conditions, any change from $\sigma_2^{D0}$ results in a decrease in the value of $v_\kappa^\infty(1) - v_\kappa^\infty(0)$, which proves that $\sigma_2^{D0}$ maximizes $\min\{v_\kappa^\infty(1) - v_\kappa^\infty(0), v_\kappa^\infty(2) - v_\kappa^\infty(0)\}$.

(ii) Since $\varepsilon > 0$ and $\alpha < 1$, we have $\eta_2(\theta) > 0$ for all $\theta = 0, 1, 2$, and thus replacing $D$ with $F$ in an element of a social strategy always improves social welfare. Hence, we first consider the social strategy $\sigma_L^F$ defined by $\sigma_L^F(\theta, \tilde{\theta}) = F$ for all $(\theta, \tilde{\theta})$. $\sigma_2^F$ maximizes social welfare $U_\kappa$ among all the social strategies with $L = 2$, but $v_\kappa^\infty(1) - v_\kappa^\infty(0) = v_\kappa^\infty(2) - v_\kappa^\infty(0) = 0$. Thus, we cannot find parameters such that $\sigma_2^F$ satisfies the incentive constraints, and thus $\sigma_2^F \notin \Gamma_2^+$. Now consider social strategies in which there is exactly one $D$ element. We can show that, under the given conditions, having $\sigma_2(\theta, \tilde{\theta}) = D$ at $(\theta, \tilde{\theta})$ such that $\tilde{\theta} > 0$ yields $v_\kappa^\infty(1) - v_\kappa^\infty(0) < 0$, whereas having $\sigma_2(\theta, \tilde{\theta}) = D$ at $(\theta, \tilde{\theta})$ such that $\tilde{\theta} = 0$ yields both $v_\kappa^\infty(1) - v_\kappa^\infty(0) > 0$ and . Thus, for any social strategy having the only $D$ element at $(\theta, \tilde{\theta})$ such that $\tilde{\theta} > 0$, there do not exist parameters in the considered parameter space with which the incentive constraint for 0-agents, $\delta(1 - 2\varepsilon)\left[v_\kappa^\infty(1) - v_\kappa^\infty(0)\right] \geq c$, is satisfied. On the other hand, for any social strategy having the only $D$ element at $(\theta, \tilde{\theta})$ such that $\tilde{\theta} = 0$, we can satisfy both incentive constraints by choosing $\beta > 0$, $\alpha < 1$, $\varepsilon < 1/2$, and $c$ sufficiently close to 0. This shows that, among the social strategies having exactly one $D$ element, only those having $D$ in the first column belong to $\Gamma_2^+$. Since $\eta_2(1) < \eta_2(0) < \eta_2(2)$, $\sigma_2^B$ achieves the highest social welfare among the three candidate social strategies. ∎

Let us try to better understand now what the conditions in Proposition 5 mean. Proposition 5(i) implies that the maximum incentive for cooperation that can be achieved with three reputations is $\beta(1 - \alpha)(1 - 2\varepsilon)b$. Hence, cooperation can be sustained with $L = 2$ if and only if $\beta(1 - \alpha)(1 - 2\varepsilon)b \geq c$. That is, if $c/b > \beta(1 - \alpha)(1 - 2\varepsilon)$, then $\sigma_2^D$ is the only feasible social strategy and thus $U_2^* = 0$. Hence, when we increase $c$ while holding other parameters fixed, we can expect that $\sigma_2^*$ changes from $\sigma_2^{D0}$ to $\sigma_2^D$ around $c = \beta(1 - \alpha)(1 - 2\varepsilon)b$. Note that the same is observed with $L = 1$ in Proposition 4. We can see that $[\eta_\tau(2)/\eta_\tau(1)][(1 - \gamma)/\gamma]$ converges to 1 as $\alpha$ goes to 0 and $\beta$ goes to 1. Hence, for given values of $b$, $c$, and $\varepsilon$, the condition (24) is satisfied and thus some cooperation can be sustained if $\alpha$ and $\beta$ are sufficiently close to 0 and 1, respectively.

Consider a social norm $\kappa = (2, \sigma_2^B)$. We obtain that

$$\min\{v_\kappa^\infty(1) - v_\kappa^\infty(0), v_\kappa^\infty(2) - v_\kappa^\infty(0)\} = v_\kappa^\infty(2) - v_\kappa^\infty(0) = (1 - \alpha)^2(1 - \varepsilon)\varepsilon(b - \beta c) \quad (25)$$

and $U_\kappa = \left[1 - (1 - \alpha)^3 (1 - \varepsilon)\varepsilon^2\right](b - c)$. Since $\Gamma_2^+$ is the set of all feasible social strategies with $L = 2$ for some parameters in the considered parameter space, we can interpret Proposition 5(ii) as stating that $\sigma_2^* = \sigma_2^B$ when the community conditions are "favorable." More precisely, we have $\sigma_2^* = \sigma_2^B$ if $(1 - \alpha)^2(1 - \varepsilon)\varepsilon(b - \beta c) \geq c$, or

$$\frac{c}{b} \leq \frac{\beta(1 - \alpha)^3(1 - 2\varepsilon)(1 - \varepsilon)\varepsilon}{1 + \beta^2(1 - \alpha)^3(1 - 2\varepsilon)(1 - \varepsilon)\varepsilon}. \quad (26)$$

Also, Proposition 5(ii) implies that $U_2^* \leq \left[1 - (1-\alpha)^3 (1-\varepsilon)\varepsilon^2\right](b-c)$ as long as the parameters remain in the considered parameter space.

In Fig 3, we show the optimal value and the optimal social strategy of $DP_2$ as we vary $c$. The optimal social strategy $\sigma_2^*$ changes in the following order before becoming $\sigma_2^D$ as $c$ increases:

$$\sigma_2^1 = \begin{bmatrix} F & F & F \\ D & F & F \\ F & F & F \end{bmatrix}, \sigma_2^2 = \begin{bmatrix} D & F & F \\ F & F & F \\ F & F & F \end{bmatrix}, \sigma_2^3 = \begin{bmatrix} D & F & F \\ D & F & F \\ F & F & F \end{bmatrix},$$

$$\sigma_2^4 = \begin{bmatrix} F & F & F \\ F & F & F \\ D & F & F \end{bmatrix}, \sigma_2^5 = \begin{bmatrix} F & F & F \\ D & F & F \\ D & F & F \end{bmatrix}, \sigma_2^6 = \begin{bmatrix} D & F & F \\ F & F & F \\ D & F & F \end{bmatrix}, \sigma_2^7 = \begin{bmatrix} D & F & F \\ D & F & F \\ D & F & F \end{bmatrix}. \quad (27)$$

Note that $\sigma_2^1 = \sigma_2^B$ for small $c$ and $\sigma_2^7 = \sigma_2^{D0}$ for large $c$ (but not too large to sustain cooperation), which are consistent with the discussion about Proposition 5. For the intermediate values of $c$, only the elements in the first column change in order to increase the incentive for cooperation. We find that the order of the optimal social strategies between $\sigma_2^1 = \sigma_2^B$ and $\sigma_2^7 = \sigma_2^{D0}$ depends on the community's parameters $(b, \beta, \alpha, \varepsilon)$.

## V. EXTENSIONS

### A. Reputation Schemes with Shorter Punishment Length

So far we have focused on maximum punishment reputation schemes under which any deviation in reported actions results in the reputation of 0. Although this class of reputation schemes is simple in that a reputation scheme can be identified with the number of reputations, it may not yield the highest social welfare among all possible reputation schemes when there are report errors. When there is no report error, i.e., $\varepsilon = 0$, an agent maintains reputation $L$ as long as it follows the prescribed social strategy. Thus, in this case, punishment exists only as a threat and it does not result in an efficiency loss. On the contrary, when $\varepsilon > 0$ and $\alpha < 1$, there exist a positive proportion of agents with reputations $0$ to $L-1$ in the stationary distribution even if all the agents follow the social strategy. Thus, there is a tension between efficiency and incentive. In order to sustain a social norm, we need to provide a strong punishment so that agents do not gain by deviation. At the same time, too severe a punishment reduces social welfare. This observation suggests that, in the presence of report errors, it is optimal to provide incentives just enough to prevent deviations. If we can provide a weaker punishment while sustaining the same social strategy, it will improve social welfare. One way to provide a weaker punishment is to use a random punishment. For example, we can consider a reputation scheme under which the reputation of a $\theta$-agent becomes 0 in the next period with probability $q_\theta \in (0, 1]$ and remains the same with probability $1 - q_\theta$ when it reportedly deviates from the social strategy. By varying the punishment probability $q_\theta$ for $\theta$-agents, we can adjust the severity of the punishment applied to $\theta$-agents. This class of reputation schemes can be identified by . Maximum punishment reputation schemes can be considered as a special case where $q_\theta = 1$ for all $\theta$.

Another way to provide a weaker punishment is to use a smaller punishment length, denoted $M$. Under the reputation scheme with $(L+1)$ reputations and punishment length $M$, reputations are updated by

$$\tau\left(\theta, \tilde{\theta}, a_R\right) = \begin{cases} \min\{\theta+1, L\} & \text{if } a_R = \sigma\left(\theta, \tilde{\theta}\right), \\ \max\{\theta-M, 0\} & \text{if } a_R \neq \sigma\left(\theta, \tilde{\theta}\right). \end{cases} \quad (28)$$

When a $\theta$-agent reportedly deviates from the social strategy, its reputation is reduced by $M$ in the next period if $\theta \geq M$ and becomes 0 otherwise. Note that this reputation scheme is analogous to real-world reputation schemes for credit rating and auto insurance risk rating. This class of reputation schemes can be identified by $(L, M)$ with $1 \leq M \leq L$.[4] Maximum punishment reputation schemes can be considered as a special case where $M = L$.

In this paper, we focus on the second approach to investigate the impacts of the punishment length on the social welfare $U_\kappa$ and the incentive for cooperation $\min_\theta \left\{ \delta(1-2\varepsilon)\left[ v_\kappa^\infty(\min\{\theta+1, L\}) - v_\kappa^\infty(\max\{\theta-M, 0\}) \right] \right\}$ of a social norm $\kappa$, which is now defined as $(L, M, \sigma)$. The punishment length $M$ affects the evolution of the reputation distribution, and the stationary distribution of reputations with the reputation scheme $(L, M)$, $\{\eta_{(L,M)}(\theta)\}_{\theta=1}^L$, satisfies the following equations:

$$\begin{aligned}
\eta_{(L,M)}(0) &= (1-\alpha)\varepsilon \sum_{\theta=0}^M \eta_{(L,M)}(\theta), \\
\eta_{(L,M)}(\theta) &= (1-\alpha)(1-\varepsilon)\eta_{(L,M)}(\theta-1) + (1-\alpha)\varepsilon\eta_{(L,M)}(\theta+M) \quad \text{for } 1 \leq \theta \leq L-M, \\
\eta_{(L,M)}(\theta) &= (1-\alpha)(1-\varepsilon)\eta_{(L,M)}(\theta-1) \quad \text{for } L-M+1 \leq \theta \leq L-1, \\
\eta_{(L,M)}(L) &= (1-\alpha)(1-\varepsilon)\{\eta_{(L,M)}(L) + \eta_{(L,M)}(L-1)\} + \alpha.
\end{aligned} \quad (29)$$

Let $\{\mu_{(L,M)}(\theta)\}_{\theta=1}^L$ be the cumulative distribution of $\{\eta_{(L,M)}(\theta)\}_{\theta=1}^L$, i.e., $\mu_{(L,M)}(\theta) = \sum_{k=0}^\theta \eta_{(L,M)}(k)$ for $\theta = 0, \ldots, L$. Fig 4 plots the stationary distribution $\{\eta_{(L,M)}(\theta)\}_{\theta=1}^L$ and its cumulative distribution $\{\mu_{(L,M)}(\theta)\}_{\theta=1}^L$ for $L = 5$ and $M = 1, \ldots, 5$. We can see that the cumulative distribution monotonically decreases with $M$, i.e. $\mu_{(L,M_1)}(\theta) \leq \mu_{(L,M_2)}(\theta)$ for all $\theta$ if $M_1 > M_2$. This shows that, as the punishment length increases, there are more agents holding a lower reputation. As a result, when the community adopts a social strategy that treats an agent with a higher reputation better, increasing the punishment length reduces social welfare while it increases the incentive for cooperation. This trade-off is illustrated in Fig 5, which plots social welfare and the incentive for cooperation under a social norm $(3, M, \sigma_3^C)$ for $M = 1, 2, 3$, where the social strategy $\sigma_L^C$ is defined by $\sigma_L^C(\theta, \tilde{\theta}) = F$ if and only if $\tilde{\theta} \geq \theta$, for all $\theta$.

---

[4] We can further generalize this class by having the punishment length depend on the reputation. That is, when a $\theta$-agent reportedly deviates from the social strategy, its reputation is reduced to $\theta - M_\theta$ in the next period for some $M_\theta \leq \theta$.

In general, the social strategy adopted in the community is determined together with the reputation scheme in order to maximize social welfare while satisfying the incentive constraints. The design problem with variable punishment lengths can be formulated as follows. First, note that the expected period payoff of a $\theta$-agent, $v_\kappa(\theta)$, can be computed by (6), with the modification of the stationary distribution to $\{\eta_{(L,M)}(\theta)\}_{\theta=1}^{L}$. Agents' long-term payoffs can be obtained by solving (8), with the transition probabilities now given by

$$p_\kappa(\theta' \mid \theta) = \begin{cases} 1-\varepsilon & \text{if } \theta' = \min\{\theta+1, L\}, \\ \varepsilon & \text{if } \theta' = \max\{\theta - M, 0\}, \\ 0 & \text{otherwise,} \end{cases} \text{ for all } \theta \in \Theta. \quad (30)$$

Finally, the design problem can be written as

$$\begin{aligned}
\underset{(L,M,\sigma)}{\text{maximize}} \quad & U_\kappa = \sum_\theta \eta_{(L,M)}(\theta) v_\kappa(\theta) \\
\text{subject to} \quad & \delta(1-2\varepsilon)\left[v_\kappa^\infty\left(\min\{\theta+1, L\}\right) - v_\kappa^\infty\left(\max\{\theta - M, 0\}\right)\right] \geq c, \\
& \forall \theta \text{ such that } \exists \tilde{\theta} \text{ such that } \sigma(\theta, \tilde{\theta}) = F, \\
& \delta(1-2\varepsilon)\left[v_\kappa^\infty\left(\min\{\theta+1, L\}\right) - v_\kappa^\infty\left(\max\{\theta - M, 0\}\right)\right] \geq -c, \\
& \forall \theta \text{ such that } \sigma(\theta, \tilde{\theta}) = D \; \forall \tilde{\theta}.
\end{aligned} \quad (31)$$

We find the optimal social strategy given a reputation scheme $(L, M)$ for $L = 3$ and $M = 1, 2, 3$, and plot the social welfare and the incentive for cooperation of the optimal social strategies in Fig 6. Since different values of $M$ induce different optimal social strategies given the value of $L$, there are no monotonic relationships between the punishment length and social welfare as well as the incentive for cooperation, unlike in Fig 5. The optimal punishment length given $L$ can be obtained by taking the punishment length that yields the highest social welfare, which is plotted in Fig 7. We can see that, as the service cost $c$ increases, the optimal punishment length increases from 1 to 2 to 3 before cooperation becomes no longer sustainable. This result is intuitive in that larger $c$ requires a stronger incentive for cooperation, which can be achieved by having a larger punishment length.

B. Whitewash-Proof Social Norms

So far we have restricted our attention to reputation schemes where newly joining agents are endowed with the highest reputation, i.e., $K = L$, without worrying about the possibility of whitewashing. We now make the initial reputation $K$ as a choice variable of the design problem while assuming that agents can whitewash their reputations in order to obtain reputation $K$ [15]. At the end of each period, agents can decide whether to whitewash their reputations or not after observing their reputations for the next period. If an agent chooses to whitewash its reputation, then it leaves and rejoins the community with $\alpha$ fraction of agents and receives initial reputation $K$. The cost of whitewashing is denoted by $c_w \geq 0$.

The incentive constraints in the design problem (18) are aimed at preventing agents from deviating from the prescribed social strategy. In the presence of potential whitewashing attempts, we need additional incentive constraints to prevent agents from whitewashing their reputations. A social norm $\kappa$ is *whitewash-proof* if and only if $v_\kappa^\infty(K) - v_\kappa^\infty(\theta) \leq c_w$ for all $\theta = 0, \ldots, L$.[5] Note that $v_\kappa^\infty(K) - v_\kappa^\infty(\theta)$ is the gain from whitewashing for an agent whose reputation is updated as $\theta$. If $v_\kappa^\infty(K) - v_\kappa^\infty(\theta) \leq c_w$, there is no net gain from whitewashing for a $\theta$-agent. We measure the *incentive for whitewashing* under a social norm $\kappa$ by $\max_{\theta \in \Theta} \{v_\kappa^\infty(K) - v_\kappa^\infty(\theta)\}$. A social norm is more effective in preventing whitewashing as the incentive for whitewashing is smaller.

To simplify our analysis, we fix the punishment length at $M = L$ so that a reputation scheme is represented by $(L, K)$ with $0 \leq K \leq L$. Let $\{\eta_{(L,K)}(\theta)\}_{\theta=1}^{L}$ be the stationary distribution of reputations under reputation scheme $(L, K)$. Then the design problem is modified as

$$\begin{aligned}
\underset{(L,K,\sigma)}{\text{maximize}} \quad & U_\kappa = \sum_\theta \eta_{(L,K)}(\theta) v_\kappa(\theta) \\
\text{subject to} \quad & \delta(1 - 2\varepsilon)\left[v_\kappa^\infty(\min\{\theta + 1, L\}) - v_\kappa^\infty(0)\right] \geq c, \ \forall \theta \text{ such that } \exists \tilde{\theta} \text{ such that } \sigma(\theta, \tilde{\theta}) = F, \\
& \delta(1 - 2\varepsilon)\left[v_\kappa^\infty(\min\{\theta + 1, L\}) - v_\kappa^\infty(0)\right] \geq -c, \ \forall \theta \text{ such that } \sigma(\theta, \tilde{\theta}) = D \ \forall \tilde{\theta}, \\
& v_\kappa^\infty(K) - v_\kappa^\infty(\theta) \leq c_w, \ \forall \theta.
\end{aligned} \tag{32}$$

Now an optimal social norm is the one that maximizes social welfare among sustainable and whitewash-proof social norms. Note that the design problem (32) always has a feasible solution for any $c_w \geq 0$ since $(L, K, \sigma_L^D)$ is sustainable and whitewash-proof for all $(L, K)$.

**Proposition 6.** If $c_w \geq \dfrac{b + c}{1 - \beta(1 - \alpha)(1 - \varepsilon)}$, then every social norm is whitewash-proof.

*Proof*: Since $-c \leq v_\kappa(\theta) \leq b$ for all $\theta$, we have $v_\kappa(\theta) - v_\kappa(\theta') \leq b + c$ for all $\theta$ and $\theta'$. Hence, by (37),

$$v_\kappa^\infty(\theta) - v_\kappa^\infty(\theta') = \sum_{l=0}^{L-1} \gamma^l \left[v_\kappa(\min\{\theta + l, L\}) - v_\kappa(\min\{\theta' + l, L\})\right] \leq \frac{b + c}{1 - \gamma}. \tag{33}$$

Therefore, if $c_w \geq (b + c)/(1 - \gamma)$, the whitewash-proof constraint is satisfied for any choice of $(L, K, \sigma)$. ∎

Now we investigate the impacts of the initial reputation $K$ on social welfare and the incentive for whitewashing. We first consider the case where the social strategy is fixed. Fig 8 plots social welfare and the incentive for whitewashing under a social norm $(3, K, \sigma_3^C)$ for $K = 0, \ldots, 3$. We can see that larger $K$ yields higher social welfare and at the same time a larger incentive for whitewashing since new agents are treated better. Hence, there is a trade-off between efficiency and whitewash-proofness as we increase $K$ while fixing the social strategy. Next we consider the optimal social strategy given a reputation scheme $(L, K)$. Fig 9 plots social welfare and the incentive for whitewashing under the optimal social strategy for $L = 3$ and $K = 0, \ldots, 3$. We can see that giving the highest reputation to new agents ($K = 3$) yields the highest social welfare but it can prevent whitewashing only for small values of $c$. With our parameter specification, choosing $K = 3$ is optimal only for small $c$, and optimal $K$ drops to 0 for other values of $c$ with which some cooperation can be sustained. Fig 10 plots the optimal initial reputation $K^*$ as we vary the whitewashing cost $c_w$, for $c = 1, 2, 3$. As $c_w$ increases, the incentive constraints for whitewashing becomes less binding, and thus $K^*$ is non-decreasing in $c_w$. On the other hand, as $c$ increases, it becomes more difficult to sustain cooperation while the difference between $v_\kappa^\infty(0)$ and $v_\kappa^\infty\left(\min\{\theta+1, L\}\right)$ increases for all $\theta$ such that $\sigma(\theta, \tilde{\theta}) = F$ for some $\tilde{\theta}$. As a result, $K^*$ is non-increasing in $c$.

---

[5] This condition assumes that an agent can whitewash its reputation only once in its lifespan in the community. More generally, we can consider the case where an agent can whitewash its reputation multiple times. For example, an agent can use a deterministic stationary decision rule for whitewashing, which can be represented by a function $w: \Theta \to \{0, 1\}$, where $w(\theta) = 1$ (resp. $w(\theta) = 0$) means that the agent whitewashes (resp. does not whitewash) its reputation if it holds reputation $\theta$ in the next period. This will yield a different expression for the gain from whitewashing.

## VI. ILLUSTRATIVE EXAMPLE

In this section, we present numerical results to illustrate in detail the performance of optimal social norms. Unless stated otherwise, the setting of the community is as follows: the benefit per service ($b = 10$), the cost per service ($c = 1$), the discount factor ($\beta = 0.8$), the turnover rate ($\alpha = 0.1$), the report error ($\varepsilon = 0.2$), the punishment step ($M = L$), and the initial reputation ($K = L$). Since the number of all possible social strategies given a punishment length $L$ increases exponentially with $L$, it takes a long time to compute the optimal social strategy even for a moderate value of $L$. Hence, we consider social norms $\kappa = \left(L, \sigma_L^*\right)$ for $L = 1, 2, 3$.

We first compare the performances of the optimal social norm and the fixed social norm for $L = 1, 2, 3$. For each $L$, we use $(L, \sigma_L^C)$ as the example of the fixed social norm. Fig 11 illustrates the results, with the red bar representing the pareto optimal value $b - c$, i.e., the highest social welfare that can be possibly sustained by a social norm, the black bar representing the social welfare of the optimal social norm, and the yellow bar representing the social welfare of $(L, \sigma_L^C)$. As it shows, the optimal social norm $(L, \sigma_L^*)$ outperforms $(L, \sigma_L^C)$. When $c$ is small, $(L, \sigma_L^*)$ delivers higher social welfare than $(L, \sigma_L^C)$. When $c$ is sufficiently large such that no cooperation can be sustained under $(L, \sigma_L^C)$, a positive level of cooperation can still be sustained under $(L, \sigma_L^*)$.

Next, we analyze the impacts of the community's parameters on the performance of optimal social norms.

**Impact of the Discount Factor**: We discuss the impact of the discount factor $\beta$ on the performance of optimal social norms. As $\beta$ increases, an agent puts a higher weight on its future payoff relative to its instant payoff. Hence, with larger $\beta$, it is easier to sustain cooperation using future reward and punishment through a social norm. This is illustrated in Fig 12(a), which shows that social welfare is non-decreasing in $\beta$.

**Impact of the Turnover Rate**: Increasing $\alpha$ has two opposite effects on social welfare. As $\alpha$ increases, the weight on the future payoffs, $\delta = \beta(1-\alpha)$, decreases, and thus it becomes more difficult to sustain cooperation. On the other hand, as $\alpha$ increases, there are more agents holding the highest reputation given the restriction $K = L$. In general, agents with the highest reputation are treated well under optimal social strategies, which implies a positive effect of increasing $\alpha$ on social welfare. From Fig 12(b), we can see that, when $\alpha$ is large, the first effect is dominant, making cooperation not sustainable. We can also see that the second effect is dominant for the values of $\alpha$ with which cooperation can be sustained, yielding an increasing tendency of social welfare with respect to $\alpha$.

**Impact of the Report Errors**: As $\varepsilon$ increases, it becomes more difficult to sustain cooperation because reward and punishment provided by a social norm becomes more random. At the same time, larger $\varepsilon$ increases the fraction of 0-agents in the stationary distribution, which usually receive the lowest long-term payoff among all reputations. Therefore, we can expect that optimal social welfare has a non-increasing tendency with respect to $\varepsilon$, as illustrated in Fig 12(c). When $\varepsilon$ is sufficiently close to 1/2, $\sigma_L^D$ is the only sustainable social strategy and social welfare falls to $0$. On the other direction, as $\varepsilon$ approaches 0, social welfare converges to its upper bound $b-c$, regardless of the punishment length, as can be seen from Proposition 1(iii). We can also observe from Fig 12 that the regions of $\alpha$ and $\varepsilon$ where some cooperation can be sustained (i.e., $U_L^* > 0$) become wider as $L$ increases, whereas that of $\beta$ is independent of $L$.

## VII. CONCLUSIONS

In this paper, we used the idea of social norms to establish a rigorous framework for the design and analysis of a class of incentive schemes to sustain cooperation in online social communities. We derived conditions for sustainable social norms, under which no agent gains by deviating from the prescribed social strategy. We formulated the problem of designing an optimal social norm and characterized optimal social welfare and optimal social strategies given parameters. We also discussed the impacts of punishment lengths and whitewashing possibility on the design and performance of optimal social norms, identifying a trade-off between efficiency and incentives. Lastly, we presented numerical results to illustrate the impacts of the discount factor, the turnover rate, and the probability of report errors on the performance of optimal social norms. Our framework will provide a foundation of incentive schemes to be deployed in real-world communities, encouraging cooperation among anonymous, self-interested individuals.

## APPENDIX A

## PROOF OF PROPOSITION 1

(i) $U^* \geq 0$ follows by noting that $(L, \sigma_L^D)$ is feasible. Note that the objective function can be rewritten as $U_\kappa = (b-c)\sum_{\theta,\tilde{\theta}} \eta_L(\theta)\eta_L(\tilde{\theta}) I(\sigma(\theta,\tilde{\theta}) = F)$, where $I$ is an indicator function. Hence, $U_\kappa \leq b-c$ for all $\kappa$, which implies $U^* \leq b-c$.

(ii) By (8), we can express $v_\kappa^\infty(1) - v_\kappa^\infty(0)$ as

$$\begin{aligned} &v_\kappa^\infty(1) - v_\kappa^\infty(0) \\ &= v_\kappa(1) + \delta\left[(1-\varepsilon)v_\kappa^\infty(2) + \varepsilon v_\kappa^\infty(0)\right] - v_\kappa(0) - \delta\left[(1-\varepsilon)v_\kappa^\infty(1) + \varepsilon v_\kappa^\infty(0)\right] \quad (34) \\ &= v_\kappa(1) - v_\kappa(0) + \delta(1-\varepsilon)\left[v_\kappa^\infty(2) - v_\kappa^\infty(1)\right]. \end{aligned}$$

Similarly, we have

$$v_\kappa^\infty(2) - v_\kappa^\infty(1) = v_\kappa(2) - v_\kappa(1) + \delta(1-\varepsilon)[v_\kappa^\infty(3) - v_\kappa^\infty(2)],$$
$$\vdots$$
$$v_\kappa^\infty(L-1) - v_\kappa^\infty(L-2) = v_\kappa(L-1) - v_\kappa(L-2) + \delta(1-\varepsilon)[v_\kappa^\infty(L) - v_\kappa^\infty(L-1)],$$
$$v_\kappa^\infty(L) - v_\kappa^\infty(L-1) = v_\kappa(L) - v_\kappa(L-1).$$
(35)

In general, for $\theta = 1, \ldots, L$,

$$v_\kappa^\infty(\theta) - v_\kappa^\infty(\theta-1) = \sum_{l=0}^{L-\theta} \gamma^l [v_\kappa(\theta+l) - v_\kappa(\theta+l-1)], \qquad (36)$$

where we define $\gamma = \delta(1-\varepsilon)$. Thus, we obtain

$$\begin{aligned}&v_\kappa^\infty(\theta) - v_\kappa^\infty(0) \\ &= v_\kappa(\theta) - v_\kappa(0) + \gamma[v_\kappa(\theta+1) - v_\kappa(1)] + \cdots + \gamma^{L-\theta}[v_\kappa(L) - v_\kappa(L-\theta)] \\ &\quad + \gamma^{L-\theta+1}[v_\kappa(L) - v_\kappa(L-\theta+1)] + \cdots + \gamma^{L-1}[v_\kappa(L) - v_\kappa(L-1)] \\ &= \sum_{l=0}^{L-1} \gamma^l [v_\kappa(\min\{\theta+l, L\}) - v_\kappa(l)],\end{aligned} \qquad (37)$$

for $\theta = 1, \ldots, L$.

Since $-c \leq v_\kappa(\theta) \leq b$ for all $\theta$, we have $v_\kappa(\theta) - v_\kappa(\tilde{\theta}) \leq b+c$ for all $(\theta, \tilde{\theta})$. Hence, by (37),

$$v_\kappa^\infty(\theta) - v_\kappa^\infty(0) \leq \frac{1-\gamma^L}{1-\gamma}(b+c) \leq \frac{b+c}{1-\gamma} \qquad (38)$$

for all $\theta = 1, \ldots, L$, for all $\kappa = (L, \sigma)$. Therefore, if $\delta(1-2\varepsilon)[(b+c)/(1-\gamma)] < c$, or equivalently, $c/b > [\beta(1-\alpha)(1-2\varepsilon)]/[1-\beta(1-\alpha)(2-3\varepsilon)]$, then the incentive constraint (15) cannot be satisfied for any $\theta$, for any social norm $(L, \sigma)$. This implies that any social strategy $\sigma$ such that $\sigma(\theta, \tilde{\theta}) = F$ for some $(\theta, \tilde{\theta})$ is not feasible, and thus $U^* = 0$.

(iii) For any $L$, define a social strategy $\sigma_L^{D0}$ by $\sigma_L^{D0}(\theta, \tilde{\theta}) = D$ for $\tilde{\theta} = 0$ and $\sigma_L^{D0}(\theta, \tilde{\theta}) = F$ for all $\tilde{\theta} > 0$, for all $\theta$. In other words, with $\sigma_L^{D0}$ each agent declines the service request of 0-agents while providing a service to other agents. Consider a social norm $\kappa = (1, \sigma_1^{D0})$. Then $v_\kappa(0) = -\eta_1(1)c$ and $v_\kappa(1) = b - \eta_1(1)c$. Hence, $U_\kappa = [1-(1-\alpha)\varepsilon](b-c)$ and $v_\kappa^\infty(1) - v_\kappa^\infty(0) = b$, and thus the incentive constraint $\delta(1-2\varepsilon)(v_\kappa^\infty(1) - v_\kappa^\infty(0)) \geq c$ is satisfied by the hypothesis $c/b \leq \beta(1-\alpha)(1-2\varepsilon)$. Since there exists a feasible solution that achieves $U_\kappa = [1-(1-\alpha)\varepsilon](b-c)$, we have $U^* \geq [1-(1-\alpha)\varepsilon](b-c)$.

(iv) Suppose, on the contrary to the conclusion, that $U^* = b - c$. If $\alpha = 1$, then (15) cannot be satisfied for any $\theta$, for any $\kappa$, which implies $U^* = 0$. Hence, it must be the case that $\alpha < 1$. Let $\kappa^* = (L^*, \sigma^*)$ be an optimal social norm that achieves $U^* = b - c$. Since $\varepsilon > 0$ and $\alpha < 1$, $\eta_{L^*}(\theta) > 0$ for all $\theta$ by (4). Since $U^* = U_{\kappa^*} = (b-c)\sum_{\theta,\tilde{\theta}} \eta_{L^*}(\theta)\eta_{L^*}(\tilde{\theta}) I(\sigma^*(\theta,\tilde{\theta}) = F)$, $\sigma^*$ should have $\sigma^*(\theta,\tilde{\theta}) = F$ for all $(\theta,\tilde{\theta})$. However, under this social strategy, all the agents are treated equally, and thus $v_{\kappa^*}^\infty(0) = \cdots = v_{\kappa^*}^\infty(L^*)$. Then $\sigma^*$ cannot satisfy the relevant incentive constraint (15) for all $\theta$ since the left-hand side of (15) is zero, which contradicts the optimality of $(L^*, \sigma^*)$.

(v) The result can be obtained by combining (i) and (iii).

(vi) Suppose that $U^* = b - c$, and let $(L, \sigma)$ be an optimal social norm that achieves $U^* = b - c$. By (iv), we obtain $\varepsilon = 0$. Then by (4), $\eta_L(\theta) = 0$ for all $0 \leq \theta \leq L - 1$ and $\eta_L(L) = 1$. Hence, $\sigma$ should have $\sigma(L, L) = F$ in order to attain $U^* = b - c$. Since $v_\kappa(L) = b - c$ and $v_\kappa(\theta) \geq -c$ for all $0 \leq \theta \leq L - 1$, we have $v_\kappa^\infty(L) - v_\kappa^\infty(0) \leq b/(1-\gamma)$ by (37). If $\delta b/(1-\delta) < c$, then the incentive constraint for $L$-agents, $\delta[v_\kappa^\infty(L) - v_\kappa^\infty(0)] \geq c$, cannot be satisfied. Therefore, we obtain $c/b \leq \delta/(1-\delta)$. ∎

APPENDIX B

PROOF OF PROPOSITION 2

Choose an arbitrary $L$. To prove the result, we will construct a social strategy $\sigma_{L+1}$ using punishment length $L+1$ that is feasible and achieves $U_L^*$. Define $\sigma_{L+1}$ by

$$\sigma_{L+1}(\theta,\tilde{\theta}) = \begin{cases} \sigma_L^*(\theta,\tilde{\theta}) & \text{for } \theta \leq L \text{ and } \tilde{\theta} \leq L, \\ \sigma_L^*(L,\tilde{\theta}) & \text{for } \theta = L+1 \text{ and } \tilde{\theta} \leq L, \\ \sigma_L^*(\theta,L) & \text{for } \theta \leq L \text{ and } \tilde{\theta} = L+1, \\ \sigma_L^*(L,L) & \text{for } \theta = L+1 \text{ and } \tilde{\theta} = L+1. \end{cases} \quad (39)$$

Let $\kappa = (L, \sigma_L^*)$ and $\kappa' = (L+1, \sigma_{L+1})$. From (4), we have $\eta_{L+1}(\theta) = \eta_L(\theta)$ for $\theta = 0, \ldots, L-1$ and $\eta_{L+1}(L) + \eta_{L+1}(L+1) = \eta_L(L)$. Using this and (6), it is straightforward to see that $v_{\kappa'}(\theta) = v_\kappa(\theta)$ for all $\theta = 0, \ldots, L$ and $v_{\kappa'}(L+1) = v_\kappa(L)$. Hence, we have that

$$\begin{aligned} U_{\kappa'} &= \sum_{\theta=0}^{L+1} \eta_{L+1}(\theta) v_{\kappa'}(\theta) = \sum_{\theta=0}^{L-1} \eta_{L+1}(\theta) v_{\kappa'}(\theta) + \sum_{\theta=L}^{L+1} \eta_{L+1}(\theta) v_{\kappa'}(\theta) \\ &= \sum_{\theta=0}^{L-1} \eta_L(\theta) v_\kappa(\theta) + \sum_{\theta=L}^{L+1} \eta_{L+1}(\theta) v_\kappa(L) \\ &= \sum_{\theta=0}^{L-1} \eta_L(\theta) v_\kappa(\theta) + \eta_L(L) v_\kappa(L) = U_\kappa = U_L^*. \end{aligned} \quad (40)$$

Using (37), we can show that $v_\kappa^\infty(\theta) - v_\kappa^\infty(0) = v_\kappa^\infty(\theta) - v_\kappa^\infty(0)$ for all $\theta = 1, \ldots, L$ and $v_\kappa^\infty(L+1) - v_\kappa^\infty(0) = v_\kappa^\infty(L) - v_\kappa^\infty(0)$. By the definition of $\sigma_{L+1}$, the right-hand side of the relevant incentive constraint (i.e., $c$ or $-c$) for each $\theta = 0, \ldots, L$ is the same both under $\sigma_L^*$ and under $\sigma_{L+1}$. Also, under $\sigma_{L+1}$, the right-hand side of the relevant incentive constraint for $\theta = L+1$ is the same as that for $\theta = L$. Therefore, $\sigma_{L+1}$ satisfies the relevant incentive constraints for all $\theta = 0, \ldots, L+1$. ∎

APPENDIX C

PROOF OF PROPOSITION 3

To facilitate the proof, we define $u_\kappa^\infty(\theta)$ by

$$u_\kappa^\infty(\theta) = \sum_{l=0}^{\infty} \gamma^l v_\kappa\left(\min\{\theta + l, L\}\right) \quad (41)$$

for $\theta = 0, \ldots, L$. Then, by (37), we have $v_\kappa^\infty(\theta) - v_\kappa^\infty(0) = u_\kappa^\infty(\theta) - u_\kappa^\infty(0)$ for all $\theta = 1, \ldots, L$. Thus, we can use $u_\kappa^\infty(\theta) - u_\kappa^\infty(0)$ instead of $v_\kappa^\infty(\theta) - v_\kappa^\infty(0)$ in the incentive constraints of $DP_L$.

Suppose that $\sigma_L^*\left(0, \hat{\theta}\right) = F$ for some $\hat{\theta}$. Then the relevant incentive constraint for a 0-agent is $\delta(1 - 2\varepsilon)\left[u_\kappa^\infty(1) - u_\kappa^\infty(0)\right] \geq c$. Suppose that $\sigma_L^*\left(0, \bar{\theta}\right) = D$ for some $\bar{\theta} \in \{1, \ldots, L-1\}$ such that $\bar{\theta} \geq \ln\frac{c}{b} / \ln \beta$. Consider a social strategy $\sigma_L'$ defined by

$$\sigma_L'\left(\theta, \tilde{\theta}\right) = \begin{cases} \sigma_L^*\left(\theta, \tilde{\theta}\right) & \text{for } \left(\theta, \tilde{\theta}\right) \neq \left(0, \bar{\theta}\right), \\ F & \text{for } \left(\theta, \tilde{\theta}\right) = \left(0, \bar{\theta}\right). \end{cases} \quad (42)$$

That is, $\sigma_L'$ is the social strategy that differs from $\sigma_L^*$ only at $\left(0, \bar{\theta}\right)$. Let $\kappa = \left(L, \sigma_L^*\right)$ and $\kappa' = \left(L, \sigma_L'\right)$. Note that $v_{\kappa'}(0) - v_\kappa(0) = -\eta_\tau\left(\bar{\theta}\right)c < 0$ and $v_{\kappa'}\left(\bar{\theta}\right) - v_\kappa\left(\bar{\theta}\right) = \eta_\tau(0)b > 0$ since $\varepsilon > 0$ and $\alpha < 1$. Thus, $U_{\kappa'} - U_\kappa = \eta_L(0)\eta_L\left(\bar{\theta}\right)(b - c) > 0$. Also,

$$u_{\kappa'}^\infty(\theta) - u_\kappa^\infty(\theta) = \begin{cases} [v_{\kappa'}(0) - v_\kappa(0)] + \gamma^{\bar{\theta}}[v_{\kappa'}\left(\bar{\theta}\right) - v_\kappa\left(\bar{\theta}\right)] \\ \quad = (1-\alpha)^{\bar{\theta}+1}(1-\varepsilon)^{\bar{\theta}}\varepsilon[\beta^{\bar{\theta}}b - c] & \text{for } \theta = 0, \\ \gamma^{\bar{\theta}-\theta}[v_{\kappa'}\left(\bar{\theta}\right) - v_\kappa\left(\bar{\theta}\right)] & \text{for } \theta = 1, \ldots, \bar{\theta}, \\ 0 & \text{for } \theta = \bar{\theta}+1, \ldots, L. \end{cases} \quad (43)$$

Since $\bar{\theta} \geq \ln\frac{c}{b} / \ln \beta$, we have $u_{\kappa'}^\infty(0) - u_\kappa^\infty(0) \leq 0$. Thus, $u_{\kappa'}^\infty(\theta) - u_{\kappa'}^\infty(0) \geq u_\kappa^\infty(\theta) - u_\kappa^\infty(0)$ for all $\theta = 1, \ldots, L$. Since $\sigma_L^*(0, \hat{\theta}) = F$ for some $\hat{\theta}$, the relevant incentive constraint for a $\theta$-agent is the same both under $\sigma_L^*$ and under $\sigma_L'$, for all $\theta$. Hence, $\sigma_L'$ satisfies the incentive constraints of $DP_L$, which contradicts the optimality of $\sigma_L^*$. This proves that $\sigma_L^*(0, \tilde{\theta}) = F$ for all $\tilde{\theta} \geq \ln\frac{c}{b} / \ln\beta$. Similar approaches can be used to prove $\sigma_L^*(0, L) = F$, (ii), and (iii). ∎

## APPENDIX D

### FIGURES

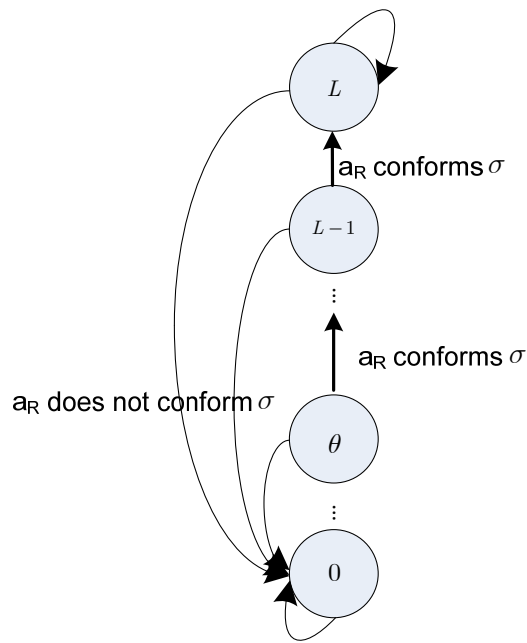

Fig 1. Schematic representation of a maximum punishment reputation scheme.

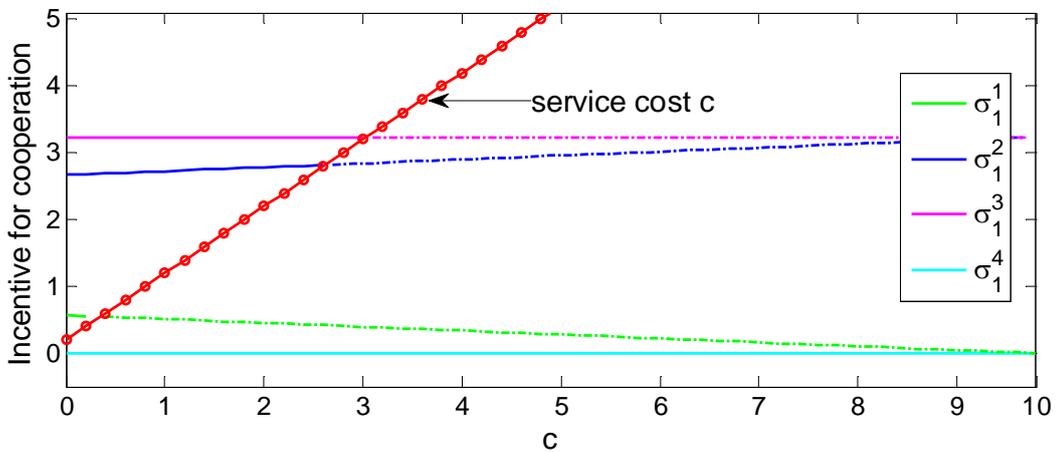

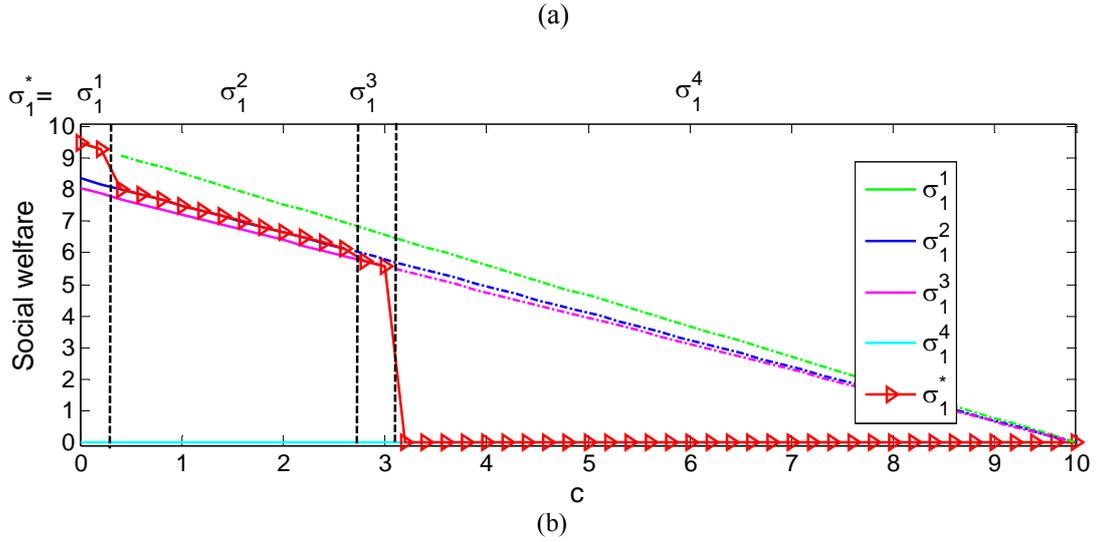

Fig 2. Performance of the four candidate social strategies when $L = 1$: (a) incentive for cooperation, and (b) social welfare and the optimal social strategy.

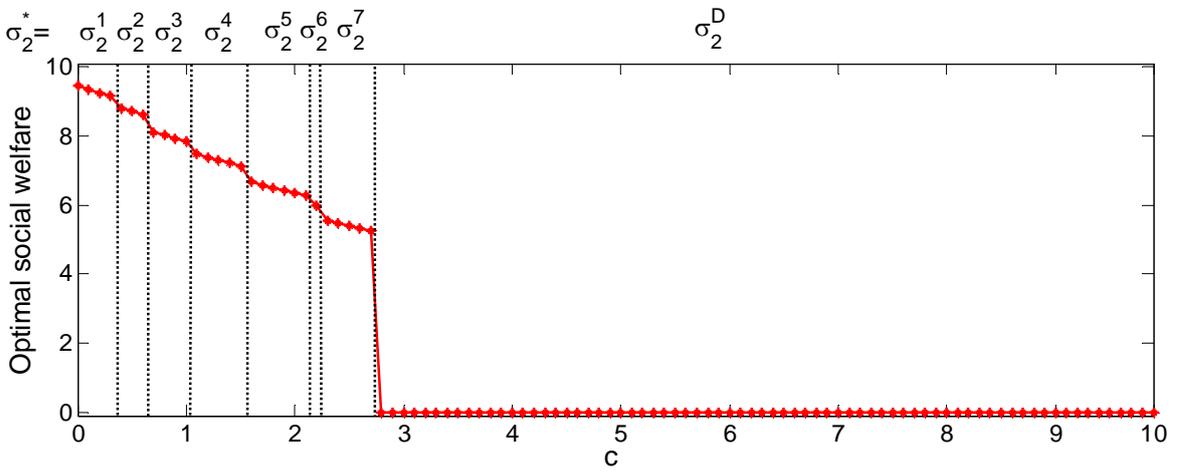

Fig 3. Optimal social welfare and the optimal social strategy of $DP_2$.

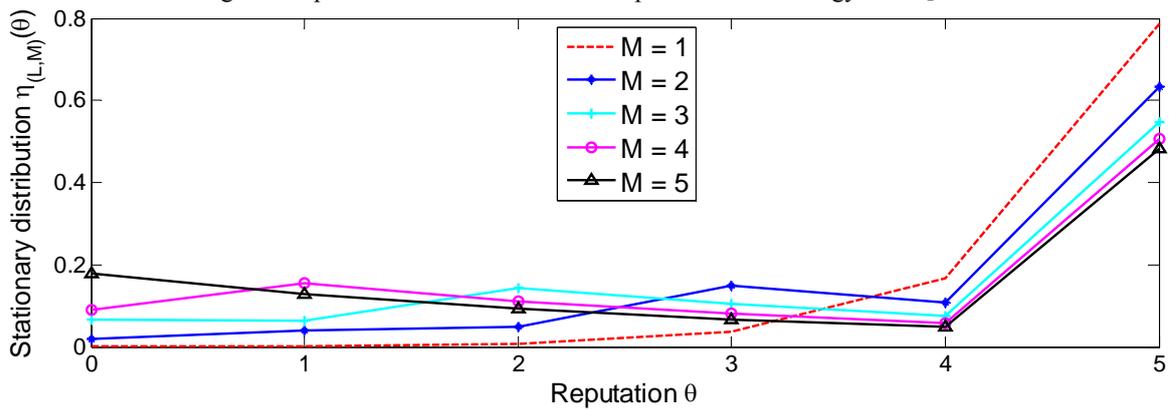

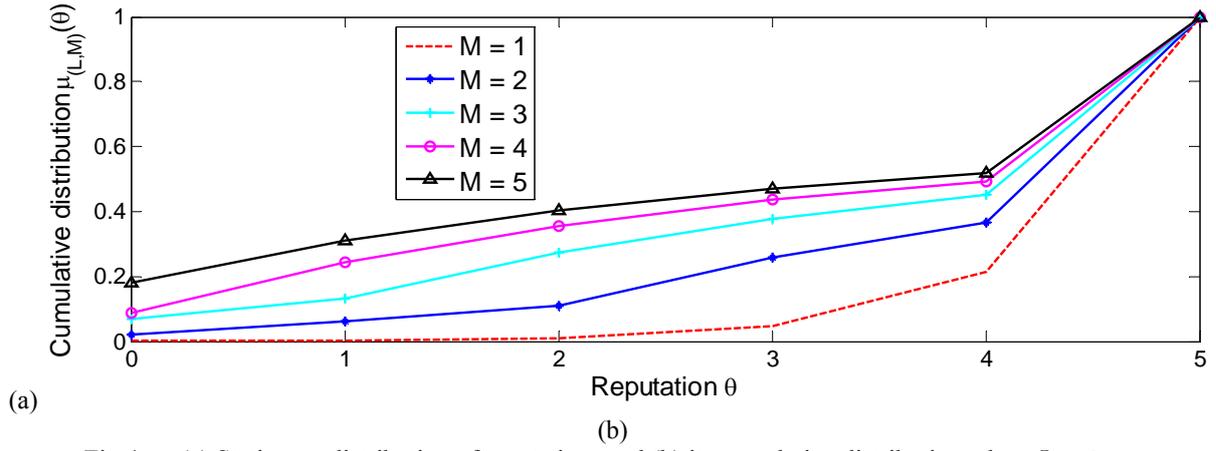

Fig 4. (a) Stationary distribution of reputations and (b) its cumulative distribution when $L = 5$.

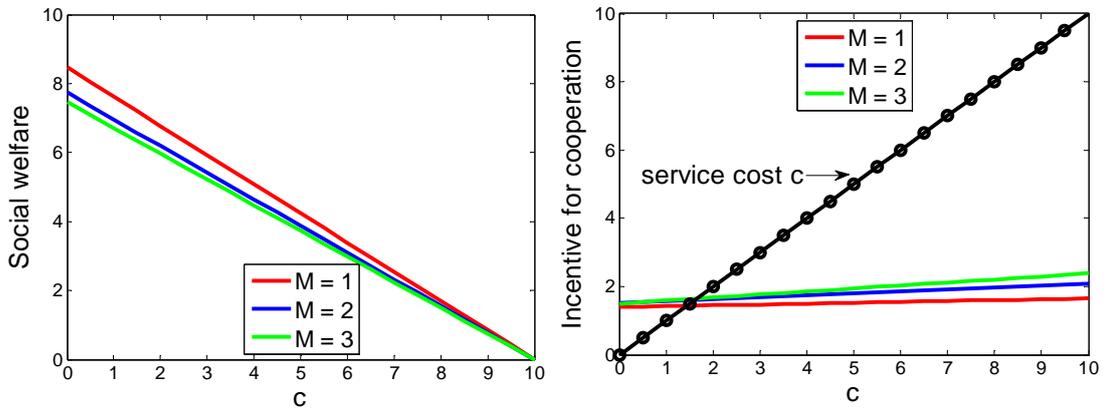

Fig 5. Social welfare and the incentive for cooperation under social strategy $\sigma_L^C$ when $L = 3$.

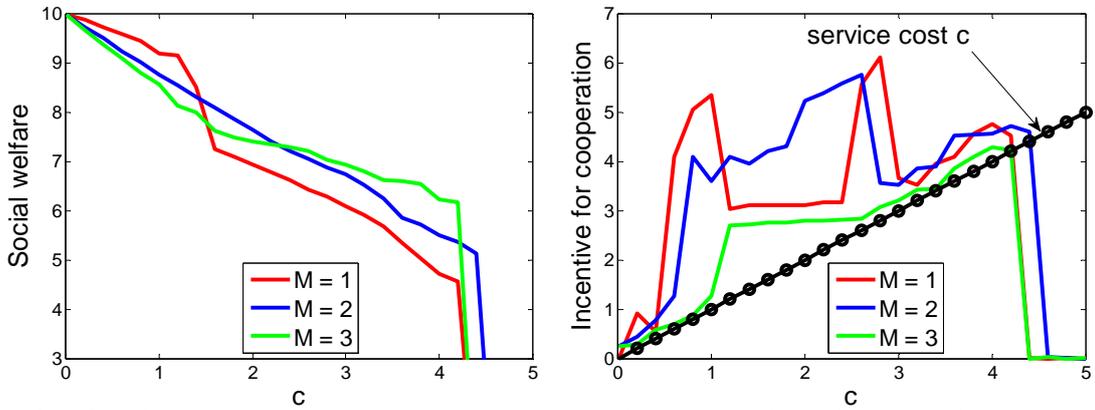

Fig 6. Social welfare and the incentive for cooperation under the optimal social strategy when $L = 3$.

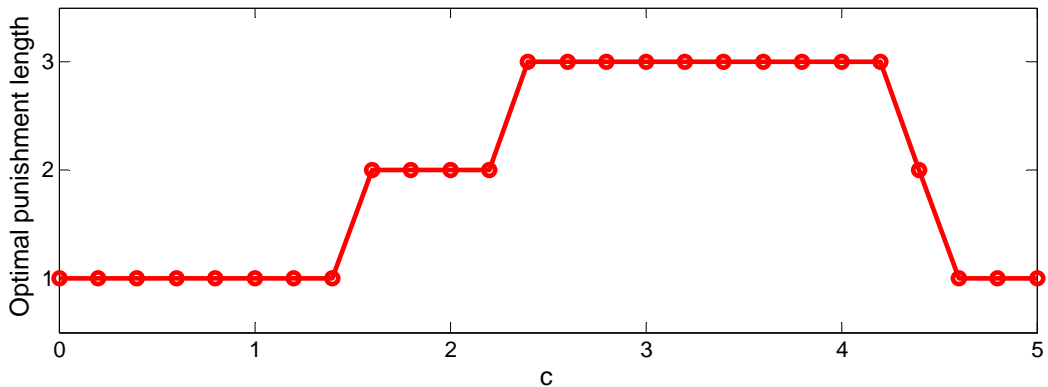

Fig 7. Optimal punishment length when $L = 3$.

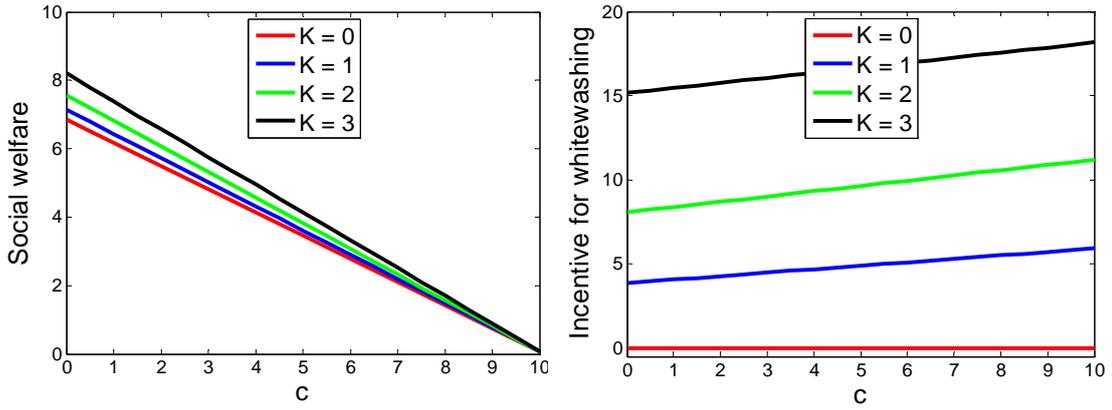

Fig 8. Social welfare and the incentive for whitewashing under social strategy $\sigma_L^C$ when $L = 3$ and $c_w = 1$.

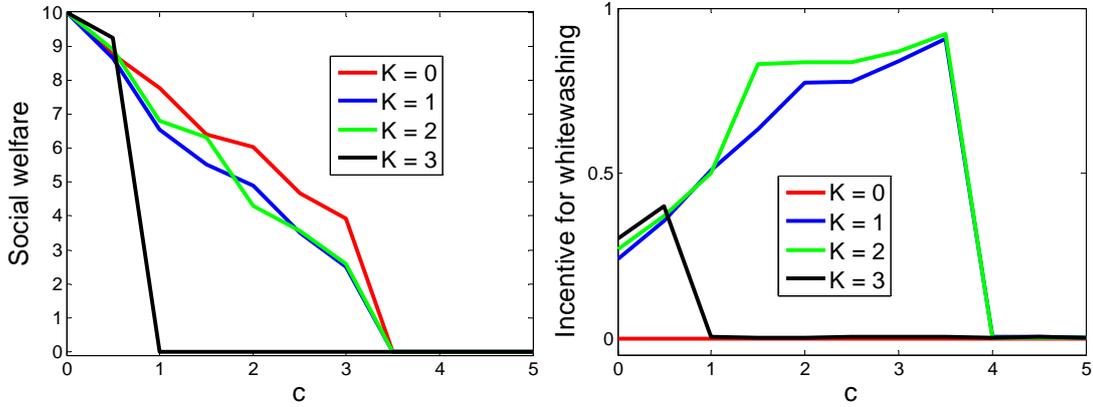

Fig 9. Social welfare and the incentive for whitewashing under the optimal social strategy when $L = 3$ and $c_w = 1$.

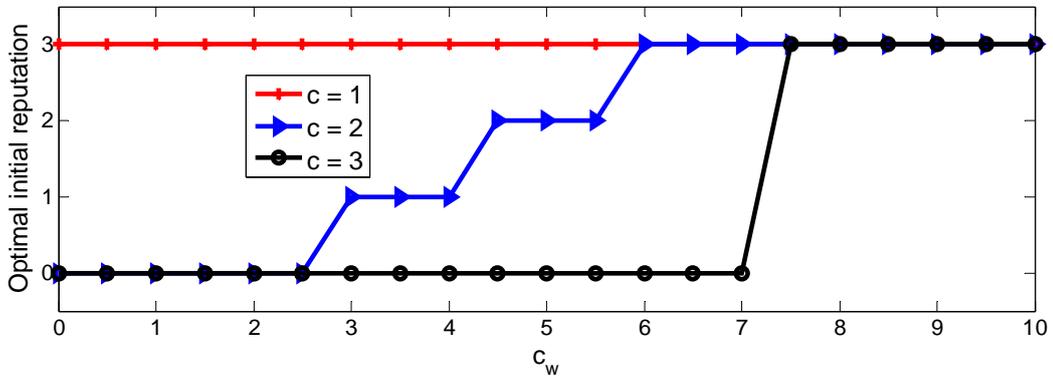

Fig 10. Optimal initial reputation when $L = 3$.

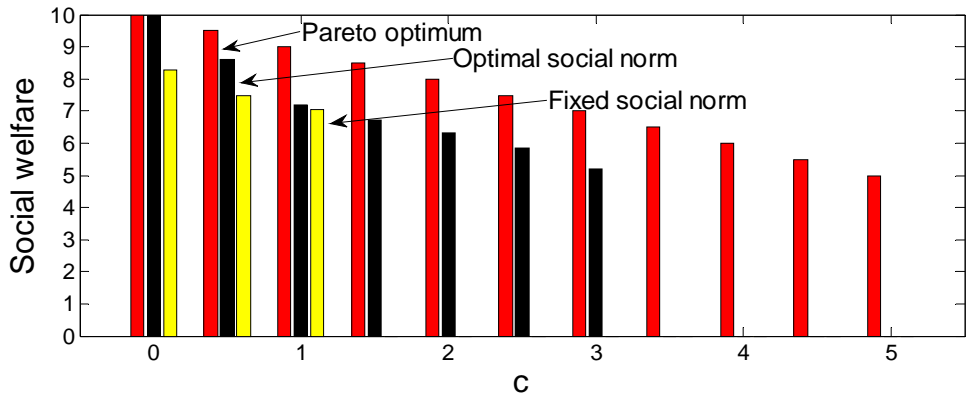

(a)

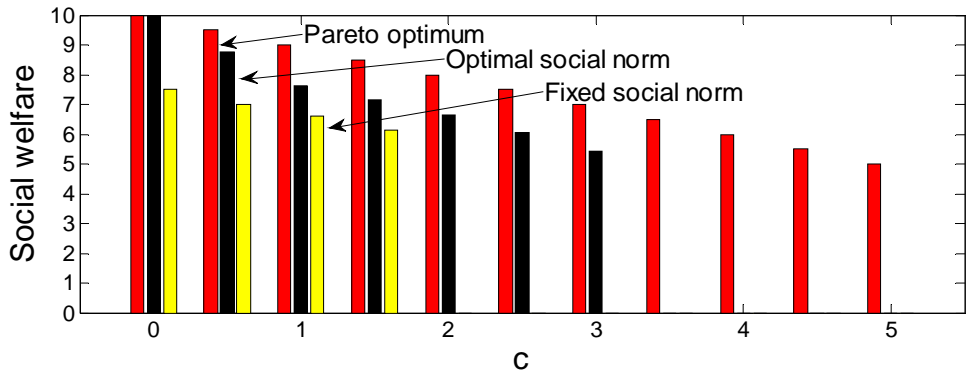

(b)

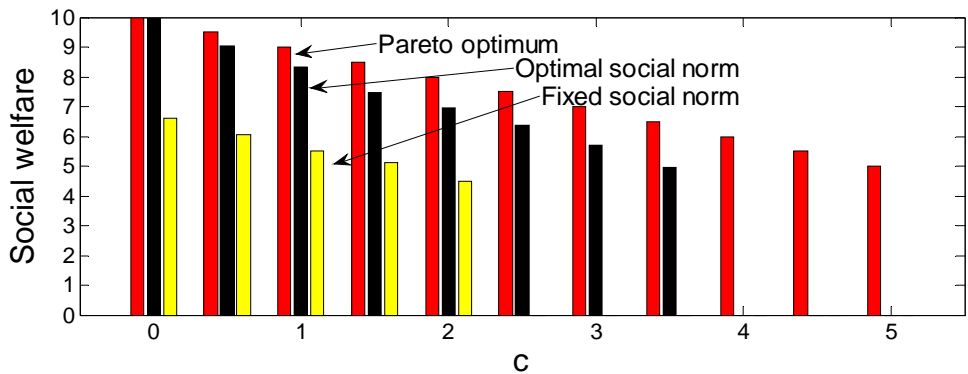

(c)

Fig 11. Performances of the optimal norm $\left(L, \sigma_L^*\right)$ and the fixed social norm $\left(L, \sigma_L^C\right)$ when (a) $L=1$; (b) $L=2$; (c) $L=3$

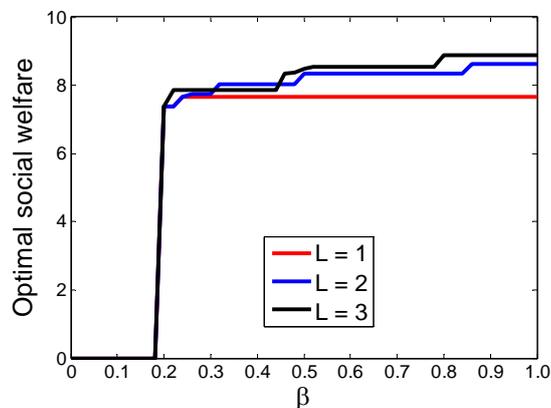

(a)

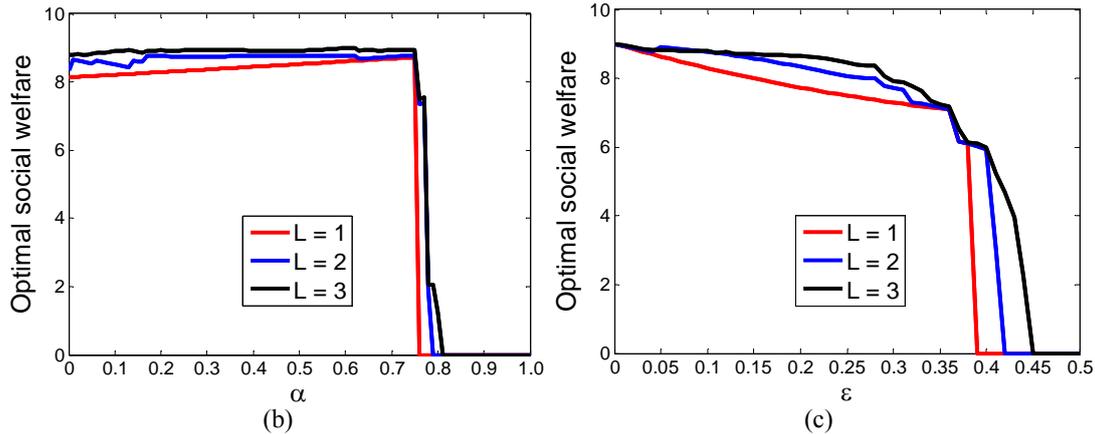

Fig 12. Optimal social welfare given *L* as $\beta$, $\alpha$, and $\varepsilon$ vary.